\documentclass[aps,10pt,prd,nofootinbib,superscriptaddress,twocolumn]{revtex4-2}
\pdfoutput=1

\usepackage[ngerman, english]{babel}
\usepackage[latin1]{inputenc}

\usepackage{mathtools}
\usepackage{amsfonts}
\usepackage{mathrsfs}
\usepackage{bbm}
\usepackage{slashed}
\usepackage{aligned-overset}
\usepackage{amssymb}

\usepackage{graphicx}
\usepackage[dvipsnames]{xcolor}
\definecolor{blue}{RGB}{31, 119, 180}
\definecolor{red}{RGB}{214, 39, 40}

\usepackage{xspace}
\usepackage{hyperref}

\usepackage{dsfont}
\usepackage{appendix}
\usepackage{enumitem}
\usepackage{booktabs}
\usepackage{units}

\newcommand{\orcid}[1]{\href{https://orcid.org/#1}{\includegraphics[height=1.9ex,width=1.9ex]{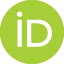}}}

\setkeys{Gin}{width=0.48\textwidth}

\graphicspath{
	{./}
}

\def\CC{{\mathcal C}}
\def\CO{{\mathcal O}}

\newcommand{\Tr}{\mathrm{Tr}}
\newcommand{\STr}{\mathrm{STr}}
\newcommand{\nn}{\nonumber }
\newcommand{\delt}{\partial_t}

\newcommand*{\overbar}[1]{\mkern 1.8mu\overline{\mkern-1.8mu#1\mkern-1.8mu}\mkern 1.8mu}
\newcommand*{\psib}{{\overbar{\psi}}}

\newcommand*{\ft}{\tilde{f}}
\newcommand*{\xt}{\tilde{x}}
\newcommand*{\Gammat}{\tilde{\Gamma}}
\newcommand*{\Rt}{\tilde{R}}

\renewcommand*{\Re}[1]{\mathfrak{Re}\left\lbrace #1\right\rbrace}
\renewcommand*{\Im}[1]{\mathfrak{Im}\left\lbrace #1\right\rbrace}
\renewcommand*\d{\mathop{}\!\mathrm{d}}
\newcommand*{\tp}{\intercal}
\newcommand*{\iu}{{\mathrm{i}\mkern1mu}}
\newcommand*{\e}{{\mathrm{e}\mkern1mu}}
\newcommand*{\dt}{\partial_t}

\newcommand*{\Nf}{N_\text{f}}
\newcommand*{\fl}{\text{fl}}
\newcommand*{\nF}{n_\text{F}}

\newcommand*{\rotate}[2]{\rotatebox[origin=c]{#1}{$#2$}}
\renewcommand{\slash}[3]{\mathrlap{\hspace*{#1 mm}\raisebox{#2 mm}[0pt][0pt]{\rotate{5}{\not}}}#3}
\newcommand*{\pvslash}{\slash{-.5}{-0.25}{\vec{p}}\vphantom{\rule[2.3mm]{0pt}{0pt}}\mkern1mu}

\newcommand{\muSB}{\mu_\text{SB}}
\newcommand*{\Res}[1]{\operatorname{Res}\left(#1\right)}
\newcommand*{\bfdot}{\makebox[1ex]{\textbf{$\cdot$}}}
\newcommand*{\stl}{\text{st}}
\newcommand*{\pll}{\text{pl}}
\newcommand*{\static}{\lim_{ Q \to 0}^{(\stl)} \mkern1mu}
\newcommand*{\plasmon}{\lim_{ Q \to 0}^{(\pll)} \mkern1mu}

\hypersetup{
	colorlinks,
	linkcolor={red!75!black},
	citecolor={blue!75!black},
	urlcolor={blue!75!black},
	pdftitle={},
	pdfauthor={Braun, Geissel, Toepfel},
	pdfkeywords={Renormalization group} {Cutoff}
	{Correlations functions} {Initial conditions},
	bookmarksopen=true,
	bookmarksopenlevel=2,
	bookmarksnumbered=true
}

\begin{document}

\title{Subtleties in the calculation of correlation functions for hot and dense systems}

\author{Sebastian T\"{o}pfel \orcid{0000-0002-0459-4382}\,} 
\affiliation{Institut f\"ur Kernphysik, Technische Universit\"at Darmstadt, 
	D-64289 Darmstadt, Germany}
\affiliation{Helmholtz Research Academy Hesse for FAIR, Campus Darmstadt, D-64289 Darmstadt, Germany}
\author{Andreas Gei\ss el \orcid{0009-0007-9283-4211}\,} 
\affiliation{Institut f\"ur Kernphysik, Technische Universit\"at Darmstadt, 
D-64289 Darmstadt, Germany}
\author{Jens Braun \orcid{0000-0003-4655-9072}\,}
\affiliation{Institut f\"ur Kernphysik, Technische Universit\"at Darmstadt, 
D-64289 Darmstadt, Germany}
\affiliation{Helmholtz Research Academy Hesse for FAIR, Campus Darmstadt, D-64289 Darmstadt, Germany}
\affiliation{ExtreMe Matter Institute EMMI, GSI, Planckstra{\ss}e 1, D-64291 Darmstadt, Germany}

\begin{abstract}
We discuss subtleties in the calculation of loop integrals in studies of hot and dense systems as they appear in both perturbative and non-perturbative approaches.
To be specific, we address subtleties which appear in situations where the order of integration, differentiation, and limit processes plays a crucial role. 
For example, this applies to computations of the effective action and the computation of momentum-dependent correlation functions.
In particular, the zero-temperature limit is delicate in systems with fermions because of the presence of discontinuities at the Fermi surface.
We provide a general discussion of scenarios where the computation and evaluation of loop integrals in the context of relativistic theories requires particular attention as a change of the order of the involved mathematical operations may lead to a different result.
Our general considerations are then illustrated with the aid of concrete examples, namely by the computation of masses from fully momentum-dependent correlation functions in the context of the Gross-Neveu-Yukawa model and quantum electrodynamics.
\end{abstract}
\maketitle
\section{Introduction}
In quantum field theory, correlation functions are the key objects under investigation since all observables can eventually be constructed from these quantities.
In particular, knowledge about $n$-point correlation functions at finite temperature and chemical potential is required for the computation of the equation of state and the phase structure of a given theory.
In general, the calculation of $n$-point functions and the derivation of physical observables from them involves integration, differentiation, and the computation of limit processes.
A priori, these mathematical operations do not commute. 
For example, interchanging a limit process with the integration over loop momenta requires certain properties from the correlation functions under consideration.
In addition, the zero-temperature limit requires attention due to the presence of discontinuities at the Fermi surface in theories with fermions at finite chemical potential.

With the present work, we aim to collect and discuss subtleties which may be encountered in calculations of $n$-point correlation functions at finite temperature and chemical potential. 
For concrete calculations, we shall employ  the functional renormalization group (fRG) approach~\cite{Wetterich:1992yh} but our general considerations are by no means bound to this approach. 
In fact, it should be added that, although the flow equation underlying the fRG approach has a one-loop structure, at least some of our considerations are readily carried over to calculations of loop integrals within other approaches.

An understanding of subtleties and complications appearing in finite-density studies is indeed of great relevance with respect to both perturbative and non-perturbative computations of the equation of state of quantum field theories as well as their phase structure. 
Our present work focuses on the careful treatment of the application of derivative operators and limiting processes to loop integrals as they arise in studies of physical systems. 
We add that we do not consider effects which emerge from changing the order of integration, which is the heart of Ref.~\cite{Gorda:2022yex}. 

The present work is organized as follows:
In Sec.~\ref{sec:fRG}, we briefly introduce the fRG approach which is mainly used in the present work for the actual derivation of loop integrals. 
Then, in Sec.~\ref{sec:Calculation_Subtleties}, we begin with a general discussion of subtleties related to interchanging the order of differentiation and integration in the computation of correlation functions, 
followed by a general consideration of the zero-temperature limit. Our discussion in this section also includes an investigation of two zero-momentum limits, namely the static and the plasmon limit.
In Sec. \ref{sec:examples}, we then illustrate our general considerations with the aid of one-loop calculations of masses from fully momentum-dependent correlation functions in the context of the Gross-Neveu-Yukawa model and quantum electrodynamics (QED).
Finally, a generalization of our considerations to functional RG flows is presented in Sec.~\ref{sec:noncomm}.
A brief summary can be found in Sec.~\ref{sec:summary}.

\section{Functional Renormalization Group Approach}
\label{sec:fRG}
Over the last decades the functional renormalization group has proven to be a valuable framework for the study of quantum field theories and statistical physics. 
It constitutes a specific implementation of the renormalization group (RG) methodology, which combines the functional approach to quantum field theory with the concept of the Wilsonian renormalization group approach. 
The fRG provides us with a non-perturbative description of the physical system which makes it especially suitable for the investigation of inherently non-perturbative phenomena such as the formation of condensates. 
In fact, the fRG approach has been successfully applied to a wide range of systems, from statistical mechanics and quantum many-particle systems over high-energy physics to gravity, see, e.g., Ref.~\cite{Dupuis:2020fhh} for a recent overview. 

The fRG is based on a characteristic evolution equation for the quantum effective average action $\Gamma_k$ with respect to an infrared (IR) scale $k$, the Wetterich equation~\cite{Wetterich:1992yh}:
\begin{align}
\label{eq:Wetterich_general}
\delt \Gamma_k[\Phi]	= \frac{1}{2} \STr \left[\left(\Gamma_k^{(1,1)}[\Phi] +R_k\right)^{-1}\cdot \delt R_k \right] 
\end{align}
with 
\begin{align}
\Gamma_k^{(1,1)}[\Phi] = \frac{\overset{\rightarrow}{\delta}}{\delta  \Phi^\tp}  \Gamma_k[\Phi] \frac{\overset{\leftarrow}{\delta}}{\delta  \Phi}\,,
\end{align}
where $t = \ln(k/\Lambda)$ denotes the dimensionless RG time and $\Lambda$ some reference scale. 
For example, we may choose~$\Lambda$ such that the effective action at this scale can be identified with the classical action, i.e., $\Gamma_{k=\Lambda}=S$. 
In all explicit calculations, we shall only consider regulator functions $R_k$ which provide an IR regularization but also render the flow ultraviolet (UV) finite.
The generalized field variable~$\Phi$ contains the degrees of freedom of the theory under consideration, e.g., gauge fields, fermions, but also composite fields. 
The supertrace runs over loop momenta, internal indices and also contains the field metric which provides a minus sign for the subspace associated with Grassmann-valued fields. 
We add that, at finite chemical potential, the trace also implies that the integration over time-like momentum modes is performed before the integration over spatial momenta.

Physical observables such as thermodynamic quantities, order parameters, and masses are in general accessible through correlations functions which can be directly obtained from the Wetterich equation~\eqref{eq:Wetterich_general} by functional differentiation. 
From this, however, it follows that the RG flow of a given $n$-point function also depends on the flows of the $(n+1)$- and $(n+2)$-correlation function. 
In most cases, we are not able to solve this infinite tower of coupled differential equations such that we need to make use of some approximation scheme. 
Since the present work is not on the construction of new truncation schemes, this is irrelevant for what follows.
It only matters that the right-hand side of the Wetterich equation assumes a simple one-loop structure which makes it advantageous for general discussions.
In our discussion of complications and subtleties potentially appearing in the computation of the effective action of hot and dense systems, we shall exploit this aspect.
Since this amounts to a detailed analysis of the structure of regularized one-loop integrals, our considerations also apply to some extent to calculations of the effective action within other approaches. 
In this respect, note also that the one-loop structure of the Wetterich equation does not imply that only one-loop corrections are included in concrete calculations. 
In fact, the propagator appearing in this equation is the full propagator which in principle allows to systematically generate loop corrections of arbitrarily high orders, see, e.g., Refs.~\cite{Papenbrock:1994kf,Litim:2001ky,Geissel:2024nmx} for a discussion.

\section{Subtleties in the computation of loop integrals at finite chemical potential}
\label{sec:Calculation_Subtleties}
In the present work we are interested in a general consideration of loop integrals with internal fermion lines which are assumed to be coupled to a quark chemical potential~\cite{Bellac:2011kqa,Kapusta:2023eix}. 
The computation of such loop diagrams is particularly delicate because of the presence of discontinuities at the Fermi surface in the zero-temperature limit. 

\subsection{Introductory remarks}
In our computation of loop integrals we shall make extensive use of residue techniques such as the Cauchy residue theorem and the Matsubara formalism, see, e.g., Refs.~\cite{Matsubara:1955ws,Gross:1980br,Das:1997gg,Altland:2006si}.
In order to apply residue techniques for the evaluation of integrals, a closed integration contour in the complex plane has to be chosen. 
Since we are primarily interested in integral expressions appearing in the context of quantum field theory, it is most useful to consider an interval on the real axis and then close the contour with a semi-circle in the upper (or lower) half of the complex plane. 
In the following we agree on the convention that $C$ denotes such a contour, where the radius~$R$ of the corresponding semi-circle is taken to infinity. 
If the integral along $C$ for some meromorphic function $f$ exists, we write
\begin{align}
\label{eq: contour integration}
&\oint_C \d z\ f(z) \nn\\ 
&\coloneqq  \lim_{R \to \infty} \left( \int_{[R,R]} \d z\ f(z) + \int_{\CC^+_0(R)} \d z\ f(z) \right)\,,
\end{align} 
where
\begin{align}
\CC^+_d(r) = \{z \in \left.\mathbb{C} \right| \Im{z}>0 \wedge |z-d|=r \}
\end{align}
describes a semi-circle in the upper half of the complex plane around the point $d \in \mathbb{R}$ with radius $r >0$. 
It follows from the residue theorem that the contour integral~\eqref{eq: contour integration} equals a weighted sum of the residues of those poles of~$f$ which lie in the interior of $C$. 
In our convention, $C$ is positively oriented and winds around a pole of~$f$ only once at most. 
For a pole $\alpha \in \mathbb{C}$ of order $n \in \mathbb{N}$, the corresponding residue is given by the formula
\begin{align}
\label{eq: definition residue}
\Res{f,\alpha} = \frac{1}{(n-1)!} \lim_{z \to \alpha}\frac{\d^{n-1}}{\d z^{n-1}} \left[ (z-\alpha)^n\ f(z) \right]\,.
\end{align}
Furthermore, if $f$ is uniformly convergent to zero on~$\CC^+_0(R)$ as~$R$ tends to infinity, i.e., if
\begin{align}
\label{eq:condition_for_vanishing_int_over_semi-circle}
\forall \varphi \in (0,\pi): \quad \lim_{R \to \infty} R\ f(R \e^{\iu \varphi}) = 0\ ,
\end{align}
then the integral along the semi-circle vanishes. As a consequence, the contour integral \eqref{eq: contour integration} simplifies to 
\begin{align}
\oint_C \d z\ f(z) = \int_\mathbb{R} \d z\ f(z)\ .
\end{align}
In the following, we shall use these relations for our general considerations of loop integrals associated with hot and dense systems.

\subsection{Differentiation under the integral sign}
\label{subsec:1}
Let $Y \subset \mathbb{R}$ be an open set, $h: \mathbb{C} \times Y \to \mathbb{C}$ as well as~$g: Y \to \mathbb{C}$ differentiable functions. Furthermore, let~$h$ be analytic on~$\mathbb{C}$. 
For $n \in \mathbb{N}$, we consider the combined function 
\begin{align}
\label{eq:generic_meromorphic_function}
f(z,y) = \frac{h(z,y)}{(z-g(y))^n}\,,
\end{align}
which is singular in all points $(z,y) \in \mathbb{C}\times Y$ that satisfy~$z= g(y)$. 
Now, let $\gamma: I \to \mathbb{C}$ be a simple closed curve in the complex plane. 
Then, the function $F$ as given by the parameter integral
\begin{align}
\label{eq:Cauchy-type_integral}
F(y) = \oint_\gamma \d z\ f(z,y)
\end{align}
is differentiable on $Y \setminus U$, where $U$ is defined by 
\begin{align}
U = \{y \in \left.Y \right| \exists\, s \in I:\ \gamma(s) = g(y) \}\, .
\end{align}
Note that elements of $U$ are parametric representations of the singularity condition $z= g(y)$ for $z$ lying on the integration contour. 
The Cauchy-type integral \eqref{eq:Cauchy-type_integral} is not defined at points at which the pole of the integrand lies on the integration contour and hence $F$ is not differentiable on $U$. 
In the following we shall assume that $U= \partial U$,\footnote{For $n=1$ in Eq.~\eqref{eq:generic_meromorphic_function} our discussion can be consistently extended to cases in which this assumption fails. } meaning that $U$ contains only isolated points such that~$F$ is differentiable almost everywhere on $Y$. We define 
\begin{align}
\forall y^\ast \in U: \quad c_{y^\ast} \coloneqq \left| h^{(n-1,0)}\left( g(y^\ast),y^\ast\right) \right|\, ,
\end{align}
which is related to the boundary value of the function $F$ as $y \to y^\ast$. In particular, if $g$ passes through the contour at $y=y^\ast$, it follows from the Sokhotski-Plemelj theorem that
\begin{align}
\left| \lim_{\varepsilon \to 0^+} F(y^\ast+ \varepsilon)-\lim_{\varepsilon \to 0^+} F(y^\ast - \varepsilon) \right| = \frac{2\pi c_{y^\ast}}{(n-1)!}\, .
\end{align}
If and only if $c_{y^\ast} = 0$ for all $y^\ast \in U$, we find that 
\begin{align}
\int_Y \d y \left( \frac{\d}{\d y} F(y) - \oint_\gamma \d z\ \partial_y f(z,y) \right) = 0\, .
\end{align}
Whenever $c_{y^\ast}$ is finite, the generalized derivative with respect to $y$ acting on $F$ generates local contributions at the non-differentiable point $y^\ast$. These local terms appear in form of Dirac delta distributions and contribute to the integral over $Y$ such that differentiation and contour integration do {\it not} commute. 
 
For illustration purposes, we now consider an exemplary function $f_\tau: \mathbb{R}^{d+1} \to \mathbb{C}$ with
\begin{align}
\label{eq:example_function_A}
f_\tau(p) = \left(\frac{1}{(p_0 + \iu \mu)^2 + x^2_\tau(\vec{p}^{\,})}\right)^n\,,
\end{align}
where we assume~$\mu > 0$ without loss of generality.\footnote{This corresponds to restricting ourselves to relativistic theories which are invariant under~$\mu\to-\mu$.} 
In loop calculations, this parameter plays the role of the chemical potential.
Furthermore, $x_\tau: \mathbb{R}^d \to \mathbb{R}$ denotes some continuous function of spatial momenta~$\vec{p}$ and $p_0$ represents the zeroth component of the vector~$p = (p_0,\vec{p}^{\,})$ associated with a $(d+1)$-dimensional Euclidean spacetime. 

The function $f_\tau$ is supposed to mimic the characteristics of a typical momentum-space integrand of a loop integral associated with an $n$-point correlation function with internal fermion lines. 
The index $\tau$ corresponds to an element of a general set of real-valued parameters, such as the RG scale, a mass parameter, a homogeneous background field, and external momenta.\footnote{In anticipation of our discussion below, we add that we always consider propagators of the form as given in Eq.~\eqref{eq:example_function_A}. Similar propagators are obtained in case of diagrams with internal boson lines.}
If $\mu$ now exceeds the ``Silver-Blaze threshold"\footnote{Because of its close relation to the Silver-Blaze property of physical systems, which describes the invariance of a theory with respect to a variation of the chemical potential for chemical potentials smaller than the pole mass of the fermions (see Refs.~\mbox{\cite{Cohen:2003kd,Marko:2014hea,Khan:2015puu,Braun:2020bhy}} for detailed discussions), we refer to~$\mu_{\text{SB}}$ as the ``Silver-Blaze threshold", see also our discussion below.}
\begin{align}
\muSB(\tau) = \min_{\vec{p}} |x_\tau(\vec{p}^{\,})|\, ,
\end{align}
then the function $f_\tau$ has a pole of order~$n \in \mathbb{N}$ in~${p = p^\ast(\tau)=(0,\vec{p}^{\,\ast}(\tau))}$, where~$\vec{p}^{\,\ast}(\tau) \in \Omega$ are the real-valued roots of $1/f_\tau$ at $p_0=0$ and
\begin{align}
\label{eq:set_of_critical_spatial_momenta}
\Omega =\{ \vec{p} \in \left.\mathbb{R}^d \right| x^2_\tau(\vec{p}^{\,})-\mu^2 =0 \}\, .
\end{align}
This set may be associated with a generalized Fermi surface. 
Next, we compute the integral of the function $f_\tau$ given in Eq.~\eqref{eq:example_function_A} with respect to the time-like momentum $p_0$ over~$\mathbb{R}$. 
In actual applications, this corresponds to the integral over the zeroth component of the loop momentum.
We note that the integrand $f_\tau$ can be analytically continued in $p_0$ and, for a better comparison with Eq.~\eqref{eq:generic_meromorphic_function}, we rewrite it as follows:
\begin{align}
f_\tau(p) = \frac{h_\tau(p_0, \vec{p}^{\,})}{\left(p_0 - g_\tau(\vec{p}^{\,})\right)^n}
\end{align}
with
\begin{align}
h_\tau(p_0, \vec{p}^{\,})  &= \left(\frac{1}{p_0 + \iu (|x_\tau(\vec{p}^{\,})|+\mu)}\right)^n
\end{align}
and
\begin{align}
\label{eq: definiton of function g}
g_\tau(\vec{p}^{\,}) &= \iu (|x_\tau(\vec{p}^{\,})|-\mu)\, .
\end{align}
For the computation of the integral over~$p_0$, we employ the residue theorem and obtain
\begin{align}
F_\tau(\vec{p}^{\,}) & = \int_{\mathbb{R}} \frac{\d p_0}{2\pi}\ f_\tau(p) \nn\\
&= \frac{\iu}{(n-1)!}\ h^{(n-1,0)}(g_\tau(\vec{p}^{\,}),\vec{p}^{\,})\ \theta(\Im{g_\tau(\vec{p}^{\,})})\nn \\
&= \binom{2n-2}{n-1} \left( \frac{1}{2 |x_\tau(\vec{p}^{\,})|} \right)^{2n-1}\ \theta(|x_\tau(\vec{p}^{\,})|-\mu)\, ,
\label{eq:parameter_integral_result}
\end{align}
which is discontinuous on
\begin{align}
U = \{\vec{p} \in \left.\mathbb{R}^d \right| \Im{g_\tau(\vec{p}^{\,})} = 0 \} = \Omega\,.
\end{align}  
The appearance of the Heaviside step function $\theta$ reflects the fact that $F_{\tau}$ is not differentiable at those spatial momenta for which the pole $g_{\tau}$ hits the real axis.
\footnote{Regarding the Heaviside step function~$\theta$, we remark that we use the convention ${\theta(0) = 1/2}$ in the present work. However, note that, for $n \geq 2$ and $\vec{p} \in U$, the integral does not exist. In that case, $F_\tau(\vec{p}^{\,\ast}_\tau)$ refers only to the finite part of the underlying integral.}
Moreover, since we have
\begin{align}
c_{\vec{p}^\ast} \propto \left( \frac{1}{2 \mu} \right)^{2n-1} > 0
\end{align}
for all $\vec{p}^{\,\ast} \in U$, we observe that the generalized derivative with respect to any parameter $\tau$ does not commute with the integration over $p_0$. 
In the following, we denote the total derivative with respect to $\tau$ by $\partial_{\tau}$. 
We then obtain
\begin{align}
\label{eq:p0_deriv_interchange}
&\int_{\vec{p}} \partial_{\tau} F_{\tau}(\vec{p}^{\,}) - \int_p \partial_{\tau} f_{\tau}(p) \\
&= \binom{2n-2}{n-1} \left( \frac{1}{2 \mu} \right)^{2n-1} \underbrace{\int_{\vec{p}} \left(\partial_{\tau} |x_{\tau}(\vec{p}^{\,})|\right) \delta(|x_{\tau}(\vec{p}^{\,})|-\mu)}_{\displaystyle \sim \theta(\mu -\muSB(\tau))}\,, \nn
\end{align}
where~$\int_p = \int {\rm d}^{4}p/(2\pi)^4$ and $\int_{\vec{p}} = \int {\rm d}^{3}p/(2\pi)^3$.
Note that this observation has far-reaching consequences for the computation of $n$-point correlation functions and, in particular, for the computation of quantities from $n$-point correlation functions (e.g., wavefunction renormalization factors) since this requires projections which often involve derivatives with respect to parameters $\tau$. 
According to our analysis, such derivatives can in general not be pulled inside the integral whenever the chemical potential exceeds the Silver-Blaze threshold. 

\subsection{Zero-temperature ambiguity}
\label{subsec:2}
Let us begin by considering an analytic function~${f: \mathbb{R} \to \mathbb{C}}$ which has an asymptotic expansion at infinity such that
\begin{align}
\label{eq:asymptotic_expansion}
f(x) = \CO\left(\frac{1}{x^s}\right) \quad \text{as}\quad  x \to \infty\ \quad \text{with} \quad s > 1 .
\end{align}
In the context of high-energy physics, such a function may represent the integrand associated with a loop integral contributing to a correlation function at zero temperature with $x$ playing the role of the time-like momentum~$p_0$. 
We therefore intend to compute its integral over the real numbers.
In order to apply the Cauchy residue theorem, we perform an analytic continuation, meaning we consistently extend the original domain of the function to an open subset of $\mathbb{C}$. We are going to assume that this function is meromorphic and we let $P$ denote the set of all poles of $f$.
Since the property~\eqref{eq:asymptotic_expansion} together with analyticity implies Eq.~\eqref{eq:condition_for_vanishing_int_over_semi-circle}, the integral of $f$ can be rewritten as a sum of residues:
\begin{align}
\label{eq:generic_zero-temp_integration}
\int_\mathbb{R} \d x\, f(x) = 2\pi \iu\ \sum_{\alpha \in P} \Res{f, \alpha} \theta(\Im{\alpha})\,,
\end{align}
where~$\theta$ is the Heaviside step function.

Letting $x$ now take on only discrete values, $x = m \in \mathbb{Z}$, the integral over $x$ turns into an infinite sum. 
Its existence is guaranteed by the property \eqref{eq:asymptotic_expansion}, see Ref.~\cite{Cohen:1979}, and, since $f$ is analytic, we can apply the Matsubara formalism to determine the value of the sum. 
This technique allows us to evaluate infinite sums in a systematic fashion by applying Cauchy's residue theorem in reverse. 
The problem of calculating the value of the sum is then shifted to the problem of constructing an integrand whose sum of residues equals the original sum. 
With the aid of an exponential weighting function, a suitable integrand is readily found:
\begin{align}
\sum_{m \in \mathbb{Z}} f(m) = 2\pi \iu\ \sum_{\alpha \in P} \Res{f\ \frac{1}{\e^{-2\pi \iu \bfdot}-1},\alpha}\,.
\end{align}
Note that this formula even holds if $f$ has poles on the real axis as long as $\mathbb{Z}\cap P = \varnothing$. 

After these general comments, we now turn to a more specific scenario and consider the case
\begin{align}
x = \nu_m(\beta) = \frac{\pi}{\beta} (2m +1)
\end{align}
with $\beta > 0$. 
This situation is encountered in thermal field theory where the parameter~$\beta$ has the meaning of the inverse temperature~$T$, i.e., $\beta =1/T$. 
In this context, $\nu_m$ is called the fermionic Matsubara frequency. Its particular form ensures that the fermion fields obey antiperiodic boundary conditions in the compactified Euclidean time direction. 
For the sum over these frequencies, we then obtain
\begin{align}
\label{eq:fermionic_Matsubara_sum}
\sum_{m \in \mathbb{Z}} f(\nu_m(\beta)) = - \iu \beta\ \sum_{\alpha \in P} \Res{f\ \frac{1}{\e^{-\iu \beta \bfdot} + 1},\alpha}\, .
\end{align}
Note that the auxiliary exponential function appearing in the sum over the residues can be related to the Fermi-Dirac distribution,
\begin{align}
\nF(x) = \frac{1}{\e^x + 1}\ .
\end{align}
This distribution function shows a non-uniform convergence in the zero-temperature limit.
To be specific, let
\begin{align}
	\mathbb{M} = \{z \in \mathbb{C}: z \notin \iu \mathbb{R}\setminus \{0 \} \}\, ,
\end{align}
then
\begin{align}
\lim_{\beta \to \infty} \nF(\beta z) = 1 - \theta(\Re{z})
\end{align}
for all $z \in \mathbb{M}$.
Even more, the Fermi-Dirac distribution function loses its complex differentiability in this limit as~$\Re{z}$ is not holomorphic. 
We can nevertheless interchange derivatives and the zero-temperature limit in the sense that $\forall z \in \mathbb{M},\ \forall x \in \mathbb{R},\ \forall n \in \mathbb{N}_0$:
\begin{align}
&\lim_{\beta \to \infty} \frac{\d^n}{\d z^n} \nF(\beta z) = \left.\left[\frac{\d^n}{\d x^n} \lim_{\beta \to \infty} \nF(\beta x)\right]\right|_{x=\Re{z}}\ .
\label{eq:limT_and_dx_commutes}
\end{align}
We focus here on fermions coupled to a chemical potential. 
For bosons, the situation is different since the Bose-Einstein distribution is not defined at $z = 0$. 
In any case, infrared divergences, which may be present in the loop integral in the limit of vanishing chemical potential, are assumed to be taken care of by standard regularization prescriptions and are therefore not further discussed.

With the above considerations at hand, we now investigate the consistency between the results~\eqref{eq:generic_zero-temp_integration} and~\eqref{eq:fermionic_Matsubara_sum} in the limit $\beta \to \infty$ for a meromorphic function~$f$ with~$P \subset \iu \mathbb{M}$. 
In the context of high-energy physics this means that our further analysis is concerned with the consistency of the results for correlation functions at finite chemical potential which have been obtained by either computing the corresponding loop diagrams directly in the zero-temperature limit or by a consideration of the zero-temperature limit of the finite-temperature results for these loop integrals.
In order to allow for a meaningful comparison, we first rescale Eq.~\eqref{eq:fermionic_Matsubara_sum} by a factor of~$2\pi/\beta$ and then take the limit as $\beta$ tends to infinity:
\begin{align}
\label{eq:zero-temp_limit}
\lim_{\beta \to \infty} \frac{2\pi}{\beta}\sum_{m \in \mathbb{Z}} f(\nu_m(\beta)) 
= 2\pi \iu\ \sum_{\alpha \in P} \Res{f\ \theta(\Im{\bfdot}),\alpha}\,.
\end{align}
Since the calculation of residues for poles of order $n$ involves a $(n-1)$-th derivative, see Eq.~\eqref{eq: definition residue}, the result above is only in accordance with the result~\eqref{eq:generic_zero-temp_integration}, if all poles in the upper half of the complex plane are of order $n\! =\! 1$.\footnote{In more general situations, in which the poles $\alpha$ can depend on further parameters, see, e.g, Eq.~\eqref{eq: definiton of function g}, the results~\eqref{eq:generic_zero-temp_integration} and~\eqref{eq:zero-temp_limit} also agree, if no pole crosses the real axis for all external parameters of interest.}
For poles of higher order, the residue involves complex derivatives which are to be understood as 
\begin{align}
0 \leq k \leq n-1: \quad \left.\left[ \frac{\d^k}{\d x^k} \theta(x) \right]\right|_{x = \Im{z}}\,,
\end{align}
when acting on the Heaviside function. 
These generalized derivatives then generate terms involving Dirac delta distributions. 
As a consequence, the zero-temperature limit of a finite-temperature calculation introduces local contributions that are missing if we work at zero temperature right from the beginning. 
To be more specific, since the computation of correlation functions requires the computation of integrals over time-like and spatial momenta, terms involving Dirac delta distributions can eventually generate finite contributions to observables. 
In general, it therefore makes a difference whether the zero-temperature results have been obtained from the consideration of the zero-temperature limit of finite-temperature results or not. 

Let us now be more concrete regarding the relationship of zero- and finite-temperature results. 
To this end, we again consider the exemplary function \eqref{eq:example_function_A}, which can be written as
\begin{align}
f_{\tau}(p) = D^{n-1}_{\tau}\ \tilde{f}_{\tau}(p)\,,
\end{align}
where the differential operator $D^n_{\tau}$ is defined by
\begin{align}
D^n_{\tau} =  \frac{(-1)^n}{n!} \left(\frac{1}{\partial_\tau x^2_{\tau}(\vec{p}^{\,})}\ \partial_\tau\right)^n 
\end{align}
for all $n \in \mathbb{N}_0$ and the auxiliary function~$\tilde{f}_{\tau}(p)$ is given by
\begin{align}
\label{eq:example_function_B}
\tilde{f}_{\tau}(p) =\frac{1}{(p_0 + \iu \mu)^2 + x^2_{\tau}(\vec{p}^{\,})}\,.
\end{align}
Using the fact that temperature-independent derivatives and the Matsubara summation commute for all $T >0$, we arrive at
\begin{align}
\label{eq:zero_temp_limit_derivative}
&\lim_{\beta \to \infty} \frac{1}{\beta}\sum_{m \in \mathbb{Z}} f_{\tau}(\nu_m(\beta),\vec{p}^{\,}) \nn \\
&\quad\overset{\eqref{eq:commutation_relation_2}}{=} \lim_{\beta \to \infty}\ D^{n-1}_{\tau}\ \frac{1}{\beta}\sum_{m \in \mathbb{Z}} \tilde{f}_{\tau}(\nu_m(\beta),\vec{p}^{\,}) \nn\\
& \quad \overset{\eqref{eq:limT_and_dx_commutes}}{=} D^{n-1}_{\tau} \lim_{\beta \to \infty} \frac{1}{\beta}\sum_{m \in \mathbb{Z}} \tilde{f}_{\tau}(\nu_m(\beta),\vec{p}^{\,}) \nn \\
&\quad \,\, = D^{n-1}_{\tau} \int_{\mathbb{R}} \frac{\d p_0}{2\pi}\ \tilde{f}_{\tau}(p)\, .
\end{align}
With our considerations from Sec.~\ref{subsec:1}, we can now verify that
\begin{align}
\label{eq:integration_order_p0}
&\forall n\geq 2\ \land \mu \geq \muSB(\tau):
\nonumber\\
&\int_{\vec{p}}\ D^{n-1}_{\tau} \int_{\mathbb{R}} \frac{\d p_0}{2\pi}\ \tilde{f}_{\tau}(p)
\neq \int_p\  D^{n-1}_{\tau} \tilde{f}_{\tau}(p)\,.
\end{align}
Loosely speaking, this implies that the zero-temperature limit of a finite-temperature loop integral always comes with at least as many terms as obtained from a direct calculation in the zero-temperature limit.

For illustration, we consider the case of $n=2$. 
The zero-temperature limit of the finite-temperature computation yields 
\begin{align}
\label{eq:example_n2}
&\lim_{\beta \to \infty} \frac{1}{\beta} \sum_{m \in \mathbb{Z}} D_\tau \tilde{f}_{\tau}(p) \nn\\
&= \frac{1}{4 x^2_\tau(\vec{p}^{\,})} \left( \frac{\theta(|x_\tau(\vec{p}^{\,})|-\mu)}{|x_\tau(\vec{p}^{\,})|}  - \delta(|x_\tau(\vec{p}^{\,})|-\mu) \right)\ .
\end{align}
For $\vec{p} \notin U$, this result is consistent with Eq.\ \eqref{eq:parameter_integral_result}, as it should be. 
However, the difference in the calculations becomes apparent when the zero-temperature pole of $f_{\tau}$ lies on the contour and becomes non-integrable. 
Since this pole is screened at finite temperature (in the Matsubara sum), the zero-temperature limit allows the divergence to assume a tangible form. 
This Dirac delta distribution then leads to additional non-trivial contributions when we integrate over the spatial momenta. 
As terms involving Dirac delta distributions do not explicitly appear in calculations directly at~$T=0$, the different approaches to the zero-temperature limit at finite chemical potential differ for $n \geq 2$. 

We close this subsection by adding that, from a mathematical point of view, our findings are simply a matter of fact and ambiguities do not arise once the procedure of mapping a multivariate function onto some value is well-defined. 
Since thermal fluctuations prevent us from reaching exactly $T=0$ in experiments, it appears reasonable to extract zero-temperature observables from finite-temperature calculations. 
Leaving this phenomenological argument aside, we shall see in the Secs.~\ref{sec:examples} and~\ref{sec:noncomm} that the strategy of preferring finite-temperature computations within the fRG approach can be based on more formal grounds.

\subsection{Static and plasmon limit at zero and finite temperature}
\label{subsec:3}
Before we discuss the static and the plasmon limit in detail, we shall consider a function~$f : \mathbb{R}^2 \to \mathbb{C}$ and an accumulation point $(x_0,y_0) \in  \mathbb{R}^2$. Then, the limits
\begin{align}
\lim_{x \to x_0} \lim_{y \to y_0} f(x,y)\ \quad\text{and}\quad 
\lim_{y \to y_0} \lim_{x \to x_0} f(x,y)
\end{align}
are called iterated limits, which in general do not need to exist. 
Note that these repeated one-variable limits are distinct from the double limit
\begin{align}
\lim_{(x,y)\to (x_0,y_0)} f(x,y)\, ,
\end{align}
which is the important limiting concept in the definition of, e.g., continuity and differentiability for bivariate functions.
Under strong assumptions, one can relate these kinds of limits~\cite{Dodd1905}, even though this might not be very useful in practical calculations.
In any case, it holds that the double limit does not exist, if both iterated limits exist but do not agree.

The appearance of iterated limits is nothing new in the context of finite-temperature studies at vanishing chemical potential. 
In fact, it is known that the momentum-dependent finite-temperature self-energy~$\Pi(Q)$ associated with a boson or fermion field is discontinuous at the origin such that approaching the point of vanishing external momentum~$Q = (Q_0, \vec{Q}^{\,}) = (0,\vec{0}^{\,})$ from different directions in momentum space results in different outcomes, see, e.g., Refs.~\cite{Weldon:1983jn,Das:1997gg,Pawlowski:2017gxj} for a discussion.
Phenomenologically speaking, finite temperature introduces a preferred Lorentz frame, where the plasma of particles and antiparticles that constitutes the heat bath is at rest. 
As a consequence, Lorentz invariance is broken explicitly and the self-energy is not a function of $Q^2$ but instead a function of $Q_0^2$ and $\vec{Q}^2$, which then allows for the existence of different limits. 
In general, the prescription of taking the time-like momentum $Q_0$ to zero first is known as the static limit whereas letting the spatial momenta vanish first is called the plasmon limit:
\begin{alignat}{2}
&\text{static limit:} \quad  && \static \Pi(Q) \coloneqq \lim_{\vec{Q} \to \vec{0}}\ \lim_{\vphantom{\vec{Q}} Q_0 \to 0}\ \Pi(Q)\ ,\\
&\text{plasmon limit:} \quad  && \plasmon \Pi(Q) \coloneqq \lim_{\vphantom{\vec{Q}} Q_0 \to 0}\ \lim_{\vec{Q} \to \vec{0}}\ \Pi(Q)\ .
\end{alignat}
When applied to the momentum-dependent self-energy, the static limit provides us with the dynamically generated curvature\footnote{In terms of the effective action, the self-energy is defined as $\Pi(Q)=\Gamma^{(1,1)}(Q)-S^{(1,1)}(Q)$. Note also that, in general, the curvature mass of a particle differs from its screening mass~$m_{\text{scr}}$ as the latter is defined by a specific zero of the corresponding renormalized two-point function: $\bar{\Gamma}^{(1,1)}(Q_0=0,|\vec{Q}\,|= \iu m_{\text{scr}})=0$, see Ref.~\cite{Helmboldt:2014iya} for a detailed discussion and also Ref.~\cite{Bellac:2011kqa} for a definition of the screening mass. In contrast to that, the curvature mass is, geometrically speaking, determined by the curvature of the effective action in a specific field direction at the ground state and it can be obtained from the renormalized two-point function in the static limit.} masses for the quantum fields in the system whereas the other limit leads to so-called plasmon masses associated with the damping of oscillations in a plasma. 
The curvature and plasmon masses may agree, but in general they do not. 
Note that finite temperature leads to a discretization of the time-like direction in momentum space such that $Q_0$ is not a continuous variable anymore. 
The limit as $Q_0$ goes to zero is then to be understood in the sense of an analytic continuation. 
For the static limit, this is equivalent to setting the external Matsubara index associated with $Q_0$ to zero.

In the following we now discuss iterated limits for external momenta in the presence of a finite chemical potential.
In particular, we shall demonstrate that a disagreement between the static and plasmon limit even occurs at zero temperature but finite chemical potential. 
Note that a finite chemical potential also induces a breaking of Lorentz invariance, even in the zero-temperature limit.
Based on Eq.\ \eqref{eq:example_function_B}, we now consider the function~$\xi: \mathbb{R}^{d+1}\times \mathbb{R}^{d+1} \to \mathbb{C}$ with
\begin{align}
\label{eq:example_function_C}
\xi(p,Q) = \ft(p) \ft(p+Q)\ ,
\end{align}
where we have dropped all the indices as those are irrelevant for our discussion of iterated limits.
This function imitates the structure of the integrand associated with a one-loop diagram with two internal fermion lines only, which contributes to, e.g., the two-point function of a boson in a Yukawa-type theory, see also Fig.~\ref{fig:meson_mass} in Sec.~\ref{sec:examples} for an illustration.
For convenience, we shall again assume~$\mu >0$. 
Using the residue theorem, the computation of the integral of $\xi$ over $p_0$, which corresponds to an evaluation of a loop integral at zero temperature, can be done analytically. Assuming~$Q \neq 0$, we find
\begin{widetext}
\begin{align}
\label{eq:contour_integral_at_finite_Q}
\Xi(\vec{p},Q) := \int_{\mathbb{R}} \frac{\d p_0}{2\pi}\ \xi(p,Q) 
= -\frac{1}{2}\left( \frac{1}{|\xt_{0}|}\  \frac{\theta(|\xt_{0}|-\mu)}{\displaystyle(|\xt_{0}|-\iu Q_0)^2 - \xt^2(\vec{Q})} 
+ \frac{1}{|\xt(\vec{Q})|}\ \frac{\theta(|\xt(\vec{Q})|-\mu)}{\displaystyle(|\xt(\vec{Q})|+\iu Q_0)^2 - \xt_{0}^2} \right)
\end{align}
\end{widetext}
with
\begin{align}
\xt(\vec{Q}) \coloneqq x(\vec{p}+\vec{Q})\, , \quad
\xt_{0} \coloneqq \xt(\vec{0}) = x(\vec{p}^{\,})\,.
\end{align}
We observe that each of the two terms contributing to the function~$\Xi$ exhibits a divergent behavior in the limit of vanishing external momenta. 
To be more precise, the residues of the integrand $\xi$ each diverge at that point where the two simple poles of $\xi$ merge to a single pole of second order.
Since the isolated singularity of $\Xi$ at~$Q=0$ is not removable for~$\mu > 0$, the integral \eqref{eq:contour_integral_at_finite_Q} is discontinuous at the origin of the vector space spanned by~$Q$. In the present case, the iterated limits yield
\begin{align}
\label{eq:static_limit_example}
\static \Xi(\vec{p},Q) &= \frac{1}{4 \xt^2_{0}} \left( \frac{\theta(|\xt_{0}|-\mu)}{|\xt_0|}  - \delta(|\xt_{0}|-\mu)\right)\ ,\\[2mm]
\label{eq:plasmon_limit_example}
\plasmon \Xi(\vec{p},Q) &= \frac{1}{4 |\xt_{0}|^3}\ \theta(|\xt_{0}|-\mu)\ .
\end{align}
We observe that the static limit generates a local contribution in form of a Dirac delta distribution, which is absent in the plasmon limit. 
Overall, we find that the chemical potential significantly affects the weighting of the residues of $\xi$ such that the contour integral of $\xi$ is in general discontinuous at~$Q=0$.
We emphasize that the same reasoning also applies, if~$\xi$ had a more complicated pole structure. However, the exemplary function~\eqref{eq:example_function_C} is already sufficient to demonstrate the emergence of different iterated limits.
In addition, we would like to point out to the relationship between the two limits above, Eqs.~\eqref{eq:static_limit_example} and \eqref{eq:plasmon_limit_example}, and the results from the previous sections. For finite external momenta, the integrand~$\xi$ has two distinct first-order poles in the complex~$p_0$-plane which differ by a $Q$-dependent offset. To some degree, this may be considered a screening of the second-order pole which emerges at~$Q = 0$. Therefore, provided that~$Q \neq 0$, the integral of $\xi$ with respect to internal momenta is well-defined. After having performed this integral, different ways of removing the offset again lead to different results. In particular, when external momenta tend to zero in the static limit, we reproduce Eq.~\eqref{eq:example_n2}, whereas the plasmon limit agrees with Eq.~\eqref{eq:parameter_integral_result} for $n=2$.

The actual computation of a correlation function also requires an integration over the spatial loop momenta.
For our schematic model of a two-point function, which is sufficient for our present discussion, this means that an integration of~$\Xi$ over the spatial loop momenta is required to eventually obtain a correction to the self-energy in the zero-temperature limit:
\begin{align}
\Pi^{(T=0)}(Q) \sim \int_{\vec{p}}\ \Xi (\vec{p},Q)\, .
\end{align}
Lorentz invariance at~$\mu=0$ then implies that the self-energy and also the underlying two-point function at~${\mu >0}$ is an even function in both~$Q_0$ and~$\vec{Q}$ but it is not a function of $Q^2$. 
Because of this explicit breakdown of the Lorentz symmetry, the two iterated limits associated with letting~$Q_0$ and~$\vec{Q}$ go to zero may differ.
Note that this is very similar to the initially discussed case of finite temperature and zero chemical potential.
From these considerations at zero and finite chemical potential, we can now deduce that different iterated limits are in general expected whenever Lorentz invariance is explicitly broken by either finite temperature or finite chemical potential. 
This observation is not limited to two-point functions. Thus, different iterated limits can equally well appear for higher $n$-point functions.

We would like to point out that our results for the static and plasmon limit at zero temperature are consistent with finite-temperature calculations since every pole of $\xi$ is of order one provided that~$Q \neq 0$. 
Specifically, we have
\begin{align}
\label{eq:full_commutation_for_Q!=0}
&\lim^{(\stl/\pll)}_{Q \to 0} \int_{\mathbb{R}} \frac{\d p_0}{2\pi}\ \xi(p,Q) \nn\\
&\quad  = \lim^{(\stl/\pll)}_{Q \to 0}  \lim_{\beta \to \infty} \frac{1}{\beta}\sum_{m \in \mathbb{Z}} \xi(\nu_m(\beta),Q) \nn\\
&\quad = \lim_{\beta \to \infty} \lim^{(\stl/\pll)}_{Q \to 0} \frac{1}{\beta}\sum_{m \in \mathbb{Z}} \xi(\nu_m(\beta),Q)\ .
\end{align}
However, we emphasize that the first and second line are no longer equal when we consider cases where the integrand $\xi$ comes with poles of higher order, see our discussion in Sec.~\ref{subsec:2}. 
This case is particularly relevant for the computation of higher-order correlation functions.

Let us now consider again the iterated limits~\eqref{eq:static_limit_example} and~\eqref{eq:plasmon_limit_example} which differ at the Fermi surface, where~${\mu = |\xt_{0}|}$. 
Consequently, the double limit as $Q \to 0$ applied to the self-energy as obtained from $\Xi$ by integration out the spatial momentum modes cannot exist for~${\mu \geq \mu_{\rm SB}}$.
We stress that $\xi$ itself is continuous in~$Q=0$ for all~$p \neq (0,\vec{p}^{\,\ast})$, where~$\vec{p}^{\,\ast}$ is defined by the solution of~$\mu = |\xt_{0}|$, see Eq.~\eqref{eq:set_of_critical_spatial_momenta}. 
This means that the operations of integrating out the $p_0$ modes and taking the limit~$Q\to 0$ do not commute simply because that double limit cannot be uniquely defined outside of the integral. 
It is nevertheless possible to realize this limit after performing the integral by choosing the correct iterated limit. In particular, we observe that the plasmon limit leads to the same result as integrating $\xi$ at zero external momenta:
\begin{align}
\label{eq:zero-temp_choses_plasmon}
\plasmon\ \int_{\mathbb{R}} \frac{\d p_0}{2\pi}\ \xi(p,Q) = \int_{\mathbb{R}} \frac{\d p_0}{2\pi}\ \lim_{Q \to 0}\xi(p,Q) 
\end{align}
We emphasize that setting the external momenta to zero in $\xi$ leads to poles of higher order in the analytically continued variable $p_0$ such that the above equation does not hold for finite-temperature calculations. 
Instead, we observe that the finite-temperature computation with the limit~$Q\to 0$ taken before the integration reproduces the static limit in the sense that
\begin{align}
\label{eq:finite-temp_choses_static}
\static\ \frac{1}{\beta} \sum_{m \in \mathbb{Z}} \xi(\nu_m(\beta),Q) = \frac{1}{\beta} \sum_{m \in \mathbb{Z}}\ \lim_{Q \to 0} \xi(\nu_m(\beta),Q)\ .
\end{align}
In fact, if it is possible in the first place to take the limit of zero external momenta after performing the Matsubara sum, then the static limit is the only unique option left ensuring consistency with Eqs.~\eqref{eq:full_commutation_for_Q!=0} and~\eqref{eq:zero-temp_choses_plasmon}. 
It is worth noting that our two findings above also hold in more general scenarios, e.g., if we allow $\xi$ to have poles of higher order,
\begin{align}
\forall n_1, n_2 \in \mathbb{N}: \quad \xi(p,Q) = \ft^{n_1}(p) \ft^{n_2}(p+Q)\,.
\end{align} 
This is because the validity of Eqs.~\eqref{eq:zero-temp_choses_plasmon} and~\eqref{eq:finite-temp_choses_static} relies on how the poles of $\xi$ contribute to the integral/series {\it and} how the components of $Q$ affect the positions of poles relevant for the integration/summation. While the former is determined by the analytic properties of $\xi$, the latter is fixed by the Lorentz symmetry in the vacuum limit.
\section{Examples}
\label{sec:examples}
So far we have presented a general discussion of subtleties which may be encountered in finite-density calculations, both at zero and finite temperature. 
In the following, we shall now discuss two concrete examples for which these subtleties become relevant. 
To be specific, we shall demonstrate the relevance of our mathematical considerations by studying the momentum-dependent two-point correlation function of the boson in the Gross-Neveu-Yukawa model and the one of the photon in QED but restricted to the limit of many fermion flavors in both cases. 
We shall always consider the case of a four-dimensional Euclidean spacetime.

For the computation of the aforementioned two-point functions, we employ the fRG approach, see Sec.~\ref{sec:fRG}. 
The restriction to the limit of many-flavors is indeed beneficial as it eventually corresponds to considering these two-point functions in a one-loop approximation for which results are already available in the literature.
Moreover, the computation of one-loop diagrams allows us in passing to demonstrate that our general considerations can also be applied to other approaches. 
We shall briefly discuss our fRG setting in the subsequent section, with a brief comment on its relation to the textbook one-loop computation of the effective action, before we then consider the Gross-Neveu-Yukawa model and QED.
\subsection{Truncation: General Consideration}
\label{subsec:trunc}
For our present purposes, it is sufficient to consider a truncation which allows us to compute the effective action at the one-loop level. 
This is obtained by simply identifying the second functional derivative of~$\Gamma_k$ with the one of the classical action $S$:
\begin{align}
\Gamma_k^{(1,1)} = S^{(1,1)}\,.
\end{align}
Since we restrict ourselves to the Gross-Neveu-Yukawa model and QED in the limit of many flavors, we only have to deal with purely fermionic loops.
Therefore, the Wetterich equation \eqref{eq:Wetterich_general} reduces to the following simplified form:
\begin{align}
\label{eq:Wetterich_truncated}
\dt \Gamma^{1\text{-loop}}_k[\Phi] 
&= -\frac{1}{2}\Tr\left[ \delt \ln  \left(S_{\psi}^{(1,1)}[\Phi] +R^\psi_k\right) \right]\nn\\
&= - \frac{1}{2}\Tr\left[\left(S_{\psi}^{(1,1)}[\Phi] +R^\psi_k\right)^{-1} \cdot \delt R^\psi_k\right]
\end{align}
with
\begin{align}
S_{\psi}^{(1,1)}[\Phi] \coloneqq \frac{\overset{\rightarrow}{\delta}}{\delta  \Psi^\tp}  S[\Phi] \frac{\overset{\leftarrow}{\delta}}{\delta  \Psi}\ , \qquad \Psi= \begin{pmatrix}
\psi\\ \psib^\tp
\end{pmatrix}\, .
\end{align}
The regulator function $R^\psi_k$ regularizes the loop diagrams considered in our present work. In momentum space, this function reads
\begin{align}
R^\psi_k(p,q) = \begin{pmatrix}
0 & - \big(\Rt_k^\psi\big)^\tp(-p) \\
\Rt_k^\psi(p) & 0 
\end{pmatrix}\ (2 \pi)^4 \delta^{(4)}(p-q)\, ,
\end{align}
where the appearance of the Dirac delta distribution above indicates momentum conservation. 
From here on, we shall restrict ourselves to ``reduced regulator functions"~$\Rt^\psi_k$ which couple only to spatial momenta.
Moreover, in order to avoid an artificial regulator-induced breaking of the chiral symmetry in our studies of the Gross-Neveu-Yukawa model and QED, we relate the momentum dependence of the regulator to the inverse propagator of the free theory and cast the ``reduced regulator function"~$\Rt^\psi_k$ into the following form~\cite{Jungnickel:1995fp}:
\begin{align}
\label{eq:3d_regulator}
\Rt^\psi_k(p) = -\pvslash\ r\left(\frac{\vec{p}^{\,2}}{k^2} \right)\, ,
\end{align} 
where $r$ denotes the dimensionless regulator shape function. 
Note that this class of regulators preserves the Silver-Blaze symmetry in the presence of a finite chemical potential~\cite{Fu:2016tey, Braun:2017srn, Braun:2020bhy} but explicitly breaks Lorentz symmetry since it only couples to the spatial momentum modes. 
This should be considered as an entirely artificial breaking of Lorentz invariance. 
In contrast to that, the breakdown of Lorentz invariance as introduced by the presence of a finite chemical potential and/or temperature is not artificial but natural. 
In fact, this naturally introduced symmetry breaking vanishes in the limit of zero external parameters whereas the regulator-induced breaking of Lorentz invariance is in general still present in the IR limit,~$k\to 0$, see, e.g., Ref.~\cite{Braun:2009si} for a detailed discussion. 
We shall now leave this aspect aside since the focus of the present work is on a discussion of subtleties which may be encountered in finite-density and finite-temperature calculations.
As we will see in our studies of the Gross-Neveu-Yukawa model and QED below, these subtleties are of very general nature and are not specific to a certain regularization scheme. 
We shall therefore employ a regulator that allows for a clean and easily accessible presentation of loop integrals.
To be specific, we choose a sharp momentum cutoff as defined by 
\begin{align}
\label{eq:sharp_cutoff}
 r(x) = \lim_{b \to \infty}\sqrt{1+ \frac{1}{x^b}}-1\, 
\end{align}
for all~$x$, see, e.g., Refs.~\cite{Pawlowski:2005xe,Braun:2014wja}.
In the following our conventions for the shape functions are such that the momentum dependence hidden in the quantity~$x=\vec{p}^{\,2}/k^2$ is not displayed explicitly and the RG-scale dependence is indicated by the subscript~$k$. 
This eventually leads us to the conventional notation~$r_k=r(\vec{p}^{\,2}/k^2)$.

A  convenient way to derive flow equations for correlation functions from the Wetterich equation relies on an expansion of the latter in powers of field degrees of freedom. 
To be more specific, we parameterize the general field variable $\Phi$ as the physical ground state~$\Phi_0$ plus fluctuations~$\Phi_\fl$ about it, $\Phi = \Phi_0 + \Phi_\fl$.
In addition, we decompose the regularized inverse propagator in terms of a field independent contribution $\mathcal{P}_k^{-1}$ and a field dependent part~$\mathcal{F}$,
\begin{align}
S^{(1,1)}_{\psi}[\Phi] + R_k = \mathcal{P}_k^{-1}[\Phi_0] + \mathcal{F}[\Phi_\fl]\, .
\end{align}
As a result, the one-loop approximation of the Wetterich equation in Eq.~\eqref{eq:Wetterich_truncated} can be cast into the following form:\footnote{The expansion relies on the assumption that the order of summation and (loop) integration as implied by the trace can be interchanged. This assumption can indeed fail, see our discussion around Eqs.~\eqref{eq:expansion_at_mean-field_level} and~\eqref{eq:cnp} in Sec.~\ref{subsec: effective potential} and our discussion of Eqs.~\eqref{eq: static limit T>0} and~\eqref{eq: plasmon limit T>0} in Sec.~\ref{subsec: full momentum dependence potential}.}
\begin{align}
\label{eq:PF_expansion}
\dt \Gamma^{1\text{-loop}}_k[\Phi]	= & - \frac{1}{2} \Tr\left[\delt \ln \mathcal P_k^{-1}[\Phi_0]\right] \\
& + \sum_{n=1}^\infty \frac{(-1)^{n}}{2n} \Tr\Big[\dt \left(\mathcal{P}_k[\Phi_0] \cdot \mathcal{F}[\Phi_\fl]  \right)^n \Big]\, . \nn
\end{align}
This can be considered as an expansion of the right-hand side of the Wetterich equation in powers of the fluctuation field $\Phi_{\rm fl}$.
Flow equations for correlation functions can now be obtained by projecting this expansion of the right-hand side of the Wetterich equation onto its left-hand side, which we shall do to compute the one-loop corrections to the boson propagator in the Gross-Neveu-Yukawa model and the photon in QED in the limit of many flavors.
Note that, for convenience, we drop the superscript of the one-loop effective action in Eqs.~\eqref{eq:Wetterich_truncated} and~\eqref{eq:PF_expansion} from here on.
\subsection{Gross-Neveu-Yukawa model}
\label{subsec:GNY}
In the following, we consider the Gross-Neveu-Yukawa which describes~$N_{\rm f}$ fermion flavors coupled to a scalar boson. 
The classical action of this model in four Euclidean spacetime dimensions reads
\begin{align}
\label{eq:GNYaction}
S[\psib, \psi, \sigma] &= \int_x\, \Big\{ \psib(x) \left[ \iu \slashed{\partial} - \iu \gamma^0 \mu + \iu h \sigma(x) \right] \psi(x)  \nn\\
& \qquad\quad +  \frac{1}{2}\sigma(x) \left[-\partial^2 + m^2\right] \sigma(x) \Big\} \, ,
\end{align}
where $\int_x=\int {\rm d}^4x$, $h$ denotes the (bare) Yukawa coupling between the fermions and the boson, and~$\psi$ represents a vector composed of $N_{\rm f}$ Dirac spinors.\footnote{Note that this model has been very frequently studied in less than four spacetime dimensions, see, e.g., Refs.~\cite{Gross:1974jv,Schon:2000qy}. For a more recent series of studies of the effective potential and correlation functions of the Gross-Neveu model in less than four spacetime dimensions, we refer the reader to Refs.~\cite{Stoll:2021ori,Koenigstein:2021llr,Koenigstein:2023yzv,Koenigstein:2024cyq}. In any case, the number of spacetime dimensions is of no relevance for our present discussion.}
The boson field~$\sigma$ does not carry any internal charge, such as flavor, and therefore does not couple to the chemical potential~$\mu$ associated with the fermions. 
Note that the Gross-Neveu-Yukawa model is distinct from the Gross-Neveu model by the fact that the boson field comes with a kinetic term in the action. 
From an RG standpoint, however, this can be simply viewed as a difference in the boundary condition for the RG flow. 

We add that the action of the Gross-Neveu-Yukawa model as well as our regulator respects the Silver-Blaze symmetry. 
Because of this, the partition function of this model does not depend on the chemical potential, provided that it remains smaller than a critical value~$\muSB$. 
This critical value is determined by the pole mass of the lowest lying fermion state in the Gross-Neveu-Yukawa model. 
For~$\mu\geq \muSB$, the generating functional then becomes non-analytic.
This is directly visible in our results presented below.
For a general discussion of this aspect, we refer the reader to Refs.~\cite{Cohen:2003kd, Marko:2014hea, Fu:2016tey,Braun:2017srn,Braun:2020bhy}.

\subsubsection{Effective potential}
\label{subsec: effective potential}

Let us now make use of the truncated Wetterich equation \eqref{eq:Wetterich_truncated} and compute the effective action of our Gross-Neveu-Yukawa model in an one-loop approximation at finite chemical potential in the limit of many fermion flavors. 
This amounts to taking only purely fermionic loops into account. 
With respect to the interaction vertices in the action~\eqref{eq:GNYaction}, it then follows that there are no loop corrections to the fermionic two-point function and therefore the fermionic wavefunction renormalization remains constant. 
Moreover, in this limit, there are also no loop corrections to the Yukawa coupling.
Nevertheless, the renormalized Yukawa coupling~$\bar{h}$ in principles acquires an RG scale dependence through the running of the bosonic wavefunction renormalization since we have~$\bar{h}=h/\sqrt{Z_{\sigma}}$.
In the loop correction to the effective potential, however, this scale dependence drops out because this loop diagram only depends on the renormalization invariant quantity~$h\sigma=\bar{h}\bar{\sigma}$. Here, $\bar{\sigma}=\sqrt{Z_{\sigma}}\sigma$ denotes the renormalized field. 
In addition to the fermion mass, the Yukawa coupling~$h$ also represents an parameter of this model, at least in four spacetime dimensions. 
In this respect, we note that the cutoff~$\Lambda$ also belongs to the definition of the model in four spacetime dimensions.

Expanding the scalar field about a homogeneous background, we arrive at the following result for the scale-dependent effective action at zero temperature:
\begin{widetext}
\begin{align}
\label{eq:effective_potential}
\frac{1}{V_4} \Gamma_k(\sigma)
 &= \frac{1}{V_4} \Gamma_\Lambda(\sigma) - 2 \Nf \int_p\ \Big.\ln\left( (p_0 + \iu \mu)^2 + \vec{p}^{\,2}(1+r_{k^\prime})^2 + h^2 \sigma^2 \right)\Big|^{k^\prime = k}_{k^\prime = \Lambda}\nn\\
&= \frac{1}{V_4} \Gamma_\Lambda(\sigma) - 2 \Nf  \int_{\vec{p}}  \left.\left[\left( \sqrt{\vec{p}^{\,2}(1+r_{k^\prime})^2 + h^2 \sigma^2}- \mu\right) \theta\left( \sqrt{\vec{p}^{\,2}(1+r_{k^\prime})^2 + h^2 \sigma^2}- \mu \right)\right] \right|^{k^\prime = k}_{k^\prime = \Lambda}\,,
\end{align}
\end{widetext}
where~$V_4$ is the four-dimensional Euclidean spacetime volume. 
The scale-dependent effective potential~$U_k$ is then given by~$U_k=\Gamma_k/V_4$.
For more details on this computation, we refer to App.~\ref{app: effective potential}.
In Eq.~\eqref{eq:effective_potential}, the scale $\Lambda$ refers to the scale at which we initialize the RG flow.
We shall choose~$\Gamma_{\Lambda}/V_4  = (1/2)m^2 \sigma^2$.

The minimum~$\sigma_0$ of the effective action is directly related to the curvature mass of the fermions,~\mbox{$m_{\rm f}=h|\sigma_0|$}. 
Note that this mass is only determined by the ratio~$h^2/m^2$, as can be deduced from Eq.~\eqref{eq:effective_potential}.
Taking this into account, the boson mass can be tuned by a variation of the Yukawa coupling~$h$ while keeping the fermion mass fixed.
This can be deduced from the expression for the curvature mass of the boson which is given by the curvature of the effective action at its minimum:
\begin{align}
m^2_\sigma  = \left.\left(\frac{{\rm d}^2}{{\rm d}\sigma^2} \frac{\Gamma (\sigma)}{V_4}\right)\right|_{\sigma=\sigma_0}\,.
\end{align}
In our one-loop approximation, we find 
\begin{widetext}
\begin{align}
m^2_\sigma 
=  \frac{1}{V_4} \Gamma^{(2)}_\Lambda\left(\frac{m_{\rm f}}{h} \right) - 2 h^2 \Nf  \int_{\vec{p}} &\Bigg[\frac{\vec{p}^{\,2}(1+r_{k^\prime})^2}{\sqrt{\vec{p}^{\,2}(1+r_{k^\prime})^2 + m_{\rm f}^2}^{\,3}}\ \theta\left(\sqrt{\vec{p}^{\,2}(1+r_{k^\prime})^2 + m_{\rm f}^2} - \mu \right) \Bigg. \nn\\
& \qquad + \Bigg. \left. \frac{m_{\rm f}^2}{\vec{p}^{\,2}(1+r_{k^\prime})^2 + m_{\rm f}^2}\ \delta\left(\sqrt{\vec{p}^{\,2}(1+r_{k^\prime})^2 + m_{\rm f}^2} - \mu \right) \right] \Bigg|_{k^\prime = \Lambda}^{k^\prime = 0}\ .
\label{eq:curvature_mass_MF_T=0}
\end{align}
\end{widetext}
The contribution from the Dirac delta distribution in Eq.~\eqref{eq:curvature_mass_MF_T=0} is relevant as it includes contributions which emerge when the chemical potential exceeds the Silver-Blaze threshold.
However, note that this term would be missing if we had taken the derivatives with respect to the field~$\sigma$ before the integration over time-like momentum modes. 
Indeed, since the effective action is non-analytic, integration and differentiation must not be interchanged. There is also another way to think about this:
If the effective action is a non-analytic function in field space, then, by definition, it is not possible to capture the full information about the system by an expansion in field degrees of freedom. 
To be more specific, we consider the expression in Eq.~\eqref{eq:effective_potential} before the integral over the momentum~$p_0$ has been computed and 
perform an expansion of it in terms of homogeneous deviations $\sigma_\fl$ from the ground state~$\sigma_0$:
\begin{align}
\label{eq:expansion_at_mean-field_level}
\int_p \ln(f(p)+h^2 \sigma^2) = \int_p\ \sum_{n=0}^\infty c_n(p) \sigma_\fl^n\ .
\end{align}
Here, $f(p)$ is a function of the four-momentum~$p$ and the chemical potential~$\mu$. The expansion coefficients~$c_n(p)$ are given by
\begin{align}
\label{eq:cnp}
&c_n(p) = \frac{1}{n!}\left.\left[\frac{\d^n}{\d \sigma_\fl^n}\ln(f(p)+h^2 (\sigma_0 + \sigma_\fl)^2) \right]\right|_{\sigma_\fl = 0}\ .
\end{align}
Summation and integration in Eq.\ \eqref{eq:expansion_at_mean-field_level} do not commute if the integral is non-analytic.\footnote{Such an interchange of summation and integration is usually made when a Taylor expansion of the effective potential is considered. Following our discussion here, this is also not allowed if the underlying loop integrals are non-analytic.}
Interchanging the order of summation and integration anyway is equivalent to applying the projection for the curvature mass directly onto the integrand and yields an incorrect result.

In order to illustrate the relevance of the contribution associated with the Dirac delta function in Eq.~\eqref{eq:curvature_mass_MF_T=0}, let us analytically evaluate this term by using the sharp cutoff in the form given in Eq.~\eqref{eq:sharp_cutoff}. 
To be specific, we compute the difference of the results as obtained by taking the second derivatives of the effective potential in Eq.~\eqref{eq:effective_potential} with respect to the field~$\sigma$ before and after the evaluation of the integral over the $p_0$ modes. 
We find
\begin{widetext}
\begin{align}
&\left.\left( \int_{\vec{p}}\ \frac{\d^2}{\d \sigma^2} \int_{\mathbb{R}} \frac{\d p_0}{2 \pi} -   \int_p\  \frac{\d^2}{\d \sigma^2} \right)  \ln\left( (p_0+ \iu \mu)^2 + \vec{p}^{\,2}(1+r_{k^\prime})^2 + h^2\sigma^2 \right)\right|_{k^\prime = \Lambda}^{k^\prime =k}\nn \\[2mm]
&=\left. h^2 \int_{\vec{p}}\ \left[ \frac{h^2\sigma^2}{\vec{p}^{\,2}(1+r_{k^\prime})^2 + h^2\sigma^2}\ \delta\left(\sqrt{\vec{p}^{\,2}(1+r_{k^\prime})^2 + h^2\sigma^2} - \mu \right) \right] \right|_{k^\prime = \Lambda}^{k^\prime =k}\nn \\[2mm]
&= \frac{h^4\sigma^2}{2\pi^2} \frac{\sqrt{\mu^2-h^2\sigma^2}}{\mu} \theta(\mu^2-h^2\sigma^2-k^2)\,.
\label{eq:bosondiff}
\end{align}
\end{widetext}
Here, we have assumed that $\Lambda$ is the largest scale which naturally restricts the range of $\mu$-values.
We note that the result above is reminiscent of Eq.~\eqref{eq:p0_deriv_interchange} for the case of $n=1$ and the Silver-Blaze threshold~\mbox{$\muSB(k, h \sigma)= \sqrt{h^2\sigma^2 + k^2}$}.
The effect of the contribution associated with the Dirac delta function on the boson curvature mass can now be obtained by evaluating Eq.~\eqref{eq:bosondiff} on the ground state in the limit~${k\to 0}$.
From this, it then follows that the Dirac delta term generates a finite contribution to the curvature mass of the boson in a regime where~$\mu \geq m_{\rm f}$, provided that the fermion mass is finite. Therefore, the chemical potential needs also to be smaller than the value~$\mu_{\text{cr}}$ associated with a potentially existing chiral phase transition above which the fermion mass is zero.
In situations with an explicit chiral symmetry breaking, the Dirac delta term gives rise to a contribution~$\sim h^2m_{\rm f}^2$ to the boson mass in the limit~$\mu/m_{\rm f}\to \infty$.

The large-$\Nf$ result for the boson curvature mass at zero temperature in Eq.~\eqref{eq:curvature_mass_MF_T=0} is perfectly consistent with the zero-temperature limit of a corresponding finite-temperature calculation, as it should be. 
To be more specific, there is no discrepancy between these results because Eq.~\eqref{eq:curvature_mass_MF_T=0} has been obtained by taking the derivatives with respect to the boson field after performing the integral over the time-like momentum modes of a function with poles in $p_0$ of order $n = 1$:
\begin{align}
m^2_\sigma  = \lim_{T \to 0} \left.\left(\frac{\d^2}{\d \sigma^2} \frac{\Gamma^{(T)} (\sigma)}{V_4} \right)\right|_{\sigma = \sigma_0}\ .
\label{eq:curvature_mass_MF_T>0}
\end{align}
Note that, on the right-hand side, the derivatives with respect to the field~$\sigma$ indeed commute with the Matsubara summation.
However, the consistency between zero-temperature and finite-temperature results can not be maintained if we take finite external momenta into account. Then, the propagator of the theory can in general not be evaluated in closed form and therefore flow equations for correlation functions would be canonically obtained by some expansion in field degrees of freedom. 
As a consequence, finite-temperature results in the zero-temperature limit will not coincide with results obtained directly at zero temperature. 
This will be shown in the following.

\subsubsection{Full momentum dependence}
\label{subsec: full momentum dependence potential}
For our discussion of the effect of finite external momenta, we introduce a generalized regularized four-momentum $\zeta_k$:
\begin{align}
\zeta_k^\tp(p) = \left(p_0 + \iu \mu,\ \vec{p}^{\,\tp} (1+r_k)\right)\, .
\label{eq:zetak}
\end{align}
Recall that the regulator shape function $r_k$ also carries a dependence on spatial momenta which we have suppressed for the sake of readabilitiy. 
We add that the use of covariant regularization schemes is in general not straightforward at finite chemical potential as a dependence of the regulator function on the chemical potential may introduce artificial poles in the propagator.

\begin{figure}[t]
	\includegraphics[width = 0.4\columnwidth]{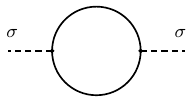}
	\caption{One-loop correction to the boson propagator in the Gross-Neveu-Yukawa model. Solid lines represent fermion propagators, dashed lines represent external boson legs.
	}
	\label{fig:meson_mass}
\end{figure}

We again take the Wetterich equation \eqref{eq:Wetterich_truncated} as our starting point for the calculation of correlation functions and perform an expansion in terms of fluctuations following Eq.\ \eqref{eq:PF_expansion}. 
For simplicity, we assume again that the ground state $\sigma_0$ is homogeneous.
The flow equation for the momentum-dependent two-point function is then obtained through functional differentiation with respect to the fluctuation field~$\sigma_{\text{fl}}$:
\begin{align}
\dt \Gamma^{(1,1)}_k\left(Q,P \right) 
&= \left( \frac{\overset{\rightarrow}{\delta}}{\delta \sigma_\fl(-Q)}\ \dt \Gamma_k\left[\sigma \right]  \frac{\overset{\leftarrow}{\delta}}{\delta \sigma_\fl(P)}\right)\Bigg|_{\sigma_\fl=0} \nn \\[2mm]
&= \dt \Gammat^{(1,1)}_k(Q)\ (2\pi)^4 \delta^{(4)}(Q-P)\, ,
\end{align}
where the flow of the reduced two-point function at zero temperature is given by 
\begin{align}
&\dt \Gammat^{(1,1)}_k(Q) \nn  \\ 
 &= -4 h^2 \Nf  \int_p\ \dt 
\frac{\zeta_k^\tp(p)\ \zeta^{\vphantom{\tp}}_k(p+Q)  - m_{\rm f}^2 }{\left( \zeta_k^2(p) + m_{\rm f}^2 \right) \left( \zeta_k^2(p+Q) + m_{\rm f}^2 \right)}\, . 
\label{eq:twopoint_PF_T=0}
\end{align}
with~$\zeta_k^2 \coloneqq \zeta_k^\tp \zeta^{\vphantom{\tp}}_k$. 
A diagrammatic representation of the right-hand side of this flow equation is given in Fig.~\ref{fig:meson_mass}. 

From Eq.~\eqref{eq:twopoint_PF_T=0}, we can now extract the curvature and plasmon mass of the boson by taking the corresponding iterated limits of vanishing external momenta.
Recall that the zero-temperature two-point function is generally non-analytic at $Q=0$ in the presence of a finite chemical potential.
As a result, the case of having zero external momentum has to be realized by an iterated limit. The static limit, where the time-like momentum~$Q_0$ is taken to zero first, is supposed to provide us with the curvature mass whereas the plasmon limit, in which the limit of vanishing spatial components~$\vec{Q}^{\,}$ is considered first, is associated with the plasmon mass. 
At zero temperature, the static and plasmon limit yield the following results:
\begin{widetext}
\begin{align}
\lim_{Q \to 0}^{(\stl)} \dt \tilde{\Gamma}^{(1,1)}_k(Q) &= -2 h^2 \Nf  \int_{\vec{p}}\ \Bigg[ \left(\dt \frac{\vec{p}^{\,2}(1+r_k)^2}{\sqrt{\vec{p}^{\,2}(1+r_k)^2+m_{\rm f}^2}^{\,3}}\right) \theta\left(\sqrt{\vec{p}^{\,2}(1+r_k)^2+m_{\rm f}^2} - \mu \right) \Bigg.\nn\\
&\qquad\qquad\qquad\qquad \Bigg.+ \left(\dt \frac{m_{\rm f}^2}{\vec{p}^{\,2}(1+r_k)^2 + m_{\rm f}^2}\right) \delta\left(\sqrt{\vec{p}^{\,2}(1+r_k)^2+m_{\rm f}^2} - \mu \right) \Bigg]\ ,
\label{eq:static_limit_T=0}\\
\lim_{Q \to 0}^{(\pll)} \dt \tilde{\Gamma}^{(1,1)}_k(Q) &= -2 h^2 \Nf  \int_{\vec{p}}\left( \dt \frac{\vec{p}^{\,2}(1+r_k)^2}{\sqrt{\vec{p}^{\,2}(1+r_k)^2+m_{\rm f}^2}^{\,3}}\right) \theta\left(\sqrt{\vec{p}^{\,2}(1+r_k)^2+m_{\rm f}^2} - \mu \right)\ .
\label{eq:plasmon_limit_T=0}
\end{align}
\end{widetext}
At first glance, the result for the static limit may appear to be in accordance with our result in Eq.~\eqref{eq:curvature_mass_MF_T=0} for the boson curvature mass as extracted from the effective potential in the large-$\Nf$ limit.
However, this is not the case as can be seen by an integration over the RG scale~$k$ in Eq.~\eqref{eq:static_limit_T=0}.
The reason for this discrepancy is that the integration over the time-like momentum modes~$p_0$ and the application of derivatives with respect to bosonic fields has been interchanged in the derivation of Eq.~\eqref{eq:static_limit_T=0}. For more details, we refer again to Eq.~\eqref{eq:expansion_at_mean-field_level} and the corresponding discussion of the interchange of summation and integration.
To be specific, the result in Eq.~\eqref{eq:static_limit_T=0} is in disagreement with the result for the boson curvature mass presented in Eq.~\eqref{eq:curvature_mass_MF_T=0}, which has been obtained by taking the field derivatives {\it after} the integration over~$p_0$.
From this, we conclude that the boson curvature mass as given in Eq.~\eqref{eq:static_limit_T=0} in the static limit is incorrect. 
The same holds for the plasmon mass in Eq.~\eqref{eq:plasmon_limit_T=0}. 

In non-perturbative calculations of correlation functions, in particular beyond the one-loop approximation, an interchange of integration and derivatives with respect to fields may appear beneficial or even necessary from a pragmatic standpoint. 
As we have seen above, this can in general be problematic. 
Still, despite this fact, it is possible to obtain correct results for correlation functions (and therefore also for the curvature and plasmon masses) in the zero-temperature limit.
Specifically, this requires to take the zero-temperature limit of the finite-temperature results for the correlation functions under consideration as we shall demonstrate in the following.

At finite temperature, the flow of the reduced two-point correlator as obtained from the expansion in Eq.~\eqref{eq:PF_expansion} reads
\begin{widetext}
\begin{align}
\dt \tilde{\Gamma}^{(1,1)(T)}_k(Q) &= -4 h^2 \Nf  \int_{\vec{p}}\ \frac{1}{\beta}\sum_{n \in \mathbb{Z}}\ \dt 
\frac{\zeta_k^\tp(\nu_n(\beta),\vec{p}^{\,})\ \zeta^{\vphantom{\tp}}_k(\nu_n(\beta) + Q_0,\vec{p} + \vec{Q}^{\,} )  - m_{\rm f}^2\  }{\Big( \zeta_k^2(\nu_n(\beta),\vec{p}^{\,}) + m_{\rm f}^2 \Big) \Big( \zeta_k^2(\nu_n(\beta) + Q_0,\vec{p} + \vec{Q}^{\,} ) + m_{\rm f}^2 \Big)}\ . \label{eq:twopoint_PF_T>0}
\end{align} 
\end{widetext}
Note already at this point that the flow of the finite-temperature two-point correlator does not reduce to the zero-temperature flow~\eqref{eq:twopoint_PF_T=0} in the limit $T \to 0$. 
As discussed on more general grounds in Sec.~\ref{subsec:2}, see especially Eqs.~\eqref{eq:zero_temp_limit_derivative} and \eqref{eq:integration_order_p0}, this discrepancy results from the fact that the integrand in Eq.~\eqref{eq:twopoint_PF_T=0} has poles of order greater than one. In particular, we remark that the derivation of flow equations for correlation functions generally includes a derivative with respect to the RG scale~$k$ which increases the order of poles in the time-like momentum variable. 
Consequently, the results for the two-point correlators computed directly at zero and finite temperature are genuinely different such that also the corresponding results for the curvature and plasmon masses differ.    
To be explicit, we find in the static and plasmon limit that
\begin{widetext}
\begin{align}
\lim_{T \to 0}\ \lim_{Q \to 0}^{(\stl)} \dt \Gammat^{(1,1)(T)}_k(Q) &= -2 h^2 \Nf \int_{\vec{p}} \dt \Bigg[ \frac{\vec{p}^{\,2}(1+r_k)^2}{\sqrt{\vec{p}^{\,2}(1+r_k)^2+m_{\rm f}^2}^3}\ \theta\left(\sqrt{\vec{p}^{\,2}(1+r_k)^2+m_{\rm f}^2} - \mu \right)\Bigg.\nn\\
&\Bigg. \qquad\qquad\qquad\qquad + \frac{m_{\rm f}^2}{\vec{p}^{\,2}(1+r_k)^2 + m_{\rm f}^2}\ \delta\left(\sqrt{\vec{p}^{\,2}(1+r_k)^2+m_{\rm f}^2} - \mu \right)\Bigg]\, ,\label{eq: static limit T>0}\\
\lim_{T \to 0}\ \lim_{Q \to 0}^{(\pll)} \dt \Gammat^{(1,1)(T)}_k(Q) &= -2 h^2 \Nf \int_{\vec{p}} \dt \Bigg[ \frac{\vec{p}^{\,2}(1+r_k)^2}{\sqrt{\vec{p}^{\,2}(1+r_k)^2+m_{\rm f}^2}^3}\ \theta\left(\sqrt{\vec{p}^{\,2}(1+r_k)^2+m_{\rm f}^2} - \mu \right) \Bigg]\, . \label{eq: plasmon limit T>0}
\end{align}
\end{widetext}
For~$\mu\geq\muSB(k, m_{\rm f})$,\footnote{In the infrared limit, $k\to 0$, we have $\muSB(0, m_{\rm f}) = m_{\rm f}$.} we observe that these results differ from those presented in Eqs.~\eqref{eq:static_limit_T=0} and~\eqref{eq:plasmon_limit_T=0}, which have been calculated directly at zero temperature. 
Moreover, Eq.\ \eqref{eq: static limit T>0} 
yields a boson curvature mass which agrees identically with the one derived from the effective potential computed directly at zero temperature,
see Eq.~\eqref{eq:curvature_mass_MF_T=0}:
\begin{align}
\dt m^2_\sigma(k) = \lim_{T \to 0}\ \lim_{Q \to 0}^{(\stl)} \dt \Gammat^{(1,1)(T)}_k(Q)\ .
\end{align}
From these results, we conclude that the momentum-dependent two-point correlator computed directly at zero temperature, which has been obtained by an interchange of integration with respect to $p_0$ and functional differentiation, does not agree with its finite-temperature pendant in the zero-temperature limit. 
Therefore, also the plasmon and curvature masses derived from these correlators are not consistent. 
As also detailed in our more general discussion in Sec.~\ref{sec:Calculation_Subtleties}, the integral with respect to $p_0$ and derivatives with respect to bosonic fields can in general not be interchanged, see Eq.~\eqref{eq:p0_deriv_interchange}, whereas an interchange of these derivatives and the Matsubara sum is allowed, see Eq.~\eqref{eq:commutation_relation_2}. 
A discussion of the full momentum dependence of correlation functions of this type will be presented elsewhere~\cite{Topfel:2024iop}. 

With respect to calculations performed directly in the zero-temperature limit, our findings indicate that an expansion of the effective action in fluctuations about the ground state is in general bound to fail for values of the chemical potential which exceed the Silver-Blaze threshold. In our model, this treshold is set by the (pole) mass of the fermions in the limit~$k\to 0$. 
Below the Silver-Blaze threshold, the partition function and also the effective action are analytic such that interchanges of integration and differentiation processes are allowed. Beyond this threshold, however, the effective action becomes non-analytic an expansion as described in Eq.\ \eqref{eq:PF_expansion} cannot correctly describe the physics of the underlying system.
Nevertheless, correct results can be obtained from such an expansion when we consider the zero-temperature limit of the finite-temperature correlation function, as explicitly demonstrated for the two-point function above.

Finally we add that our computation of $n$-point function naturally relies on an expansion of propagators in terms of fields, see Eq.~\eqref{eq:PF_expansion}.
If we nevertheless insist on a calculation of $n$-point correlation functions directly at zero temperature without ``making a detour" to finite temperature, then we have to define a hands-on prescription which allows us to include the missing terms. 
For the momentum-dependent two-point correlator, for example, we observe that it suffices to interchange the derivative with respect to the RG scale~$k$ and the integral over the time-like momentum modes~$p_0$ in order to recover the terms missing in a zero-temperature calculation.
This suggests that the correct result for loop corrections can also be obtained from calculations directly at zero temperature within the fRG formalism by suitably rewriting the loop integrals. 
To be specific, the kernel~$f_\tau$ of a loop integral associated with a given $n$-point function has to be rewritten in terms of a function~$\ft_\tau$, which has only poles of order one, and a suitably defined derivative operator $D_\tau^{n-1}$, see Sec.~\ref{subsec:2}.
Recall that the index $\tau$  denotes a placeholder for a parameter on which the propagator depends, e.g., a mass or the RG scale. 
We then find that computing the integral of $\ft_\tau$ with respect to the time-like momentum modes {\it before} applying the derivative operator~$D_\tau^{n-1}$ yields results which agree with the zero-temperature limit of a corresponding finite-temperature calculation.
Although it is in principle not allowed to interchange the order of integration and differentiation, see Sec.~\ref{subsec:1}, it becomes clear from our discussion in Secs.~\ref{subsec:2} and~\ref{sec:noncomm} why this hands-on prescription still yields correct results for correlation functions in the zero-temperature limit.

\begin{figure}[t]
	\includegraphics[width = 0.4\columnwidth]{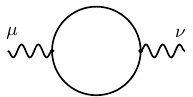}
	\caption{Feynman diagram of the one-loop correction to the photon propagator. Solid lines represent fermion propagators where wavy lines represent amputated photon legs and are associated with the Lorentz indices $\mu$ and $\nu$.}
	\label{fig:polTensor}
\end{figure}

\subsection{Quantum Electrodynamics}
\label{subsec:photon_curvature_masses}
In quantum electrodynamics, the free propagator of the gauge boson, i.e., the photon, receives corrections from electron-photon interactions. 
In the limit of many fermion flavors, which we shall consider here, this correction is of order~$\mathcal{O}(e^2)$, see Fig.~\ref{fig:polTensor}.
Because of this quantum correction, photons acquire a finite curvature and plasmon mass at finite temperature and/or chemical potential.
These masses can be extracted from the loop diagram depicted in Fig.~\ref{fig:polTensor} in the limit of vanishing external momenta.
In this section, we shall compute these masses at finite temperature and chemical potential. Moreover, we shall also consider the zero-temperature limit in the presence of a finite chemical potential.

In Euclidean spacetime, the QED action in the chiral limit reads
\begin{align}
	\hspace{-0.07cm}S[\bar{\psi},\psi,A_{\mu}]  = \int_x &\Big\lbrace  \psib(x)  \left( {\rm i} \slashed \partial + e \slashed A(x) -  {\rm i} \mu\gamma_0 \right) \psi(x) 
	\nn\\
	&+ \frac{1}{4}F_{\mu\nu}(x)F_{\mu\nu}(x)\Big\rbrace + \delta S_{\rm gf}\ ,
\end{align} 
where~$\psi$ represents a vector composed of $N_{\rm f}$ Dirac spinors,~$A_{\mu}$ is the gauge field (photon), and $e$ denotes the (bare) coupling between the fermions and the photons.
In addition to their coupling to the gauge fields, the fermions are coupled to the chemical potential~$\mu$.
Last but not least, $\delta S_{\rm gf}$ denotes the standard gauge fixing term for covariant gauges.

In this work, we are only interested in the photon polarization tensor~$\Pi_{\mu\nu}(Q)$.
In the vacuum limit, this tensor, which can be expanded in terms of the Lorentz tensors $\delta_{\mu\nu}$ and $Q_\mu Q_\nu$, is transversal, \mbox{$Q_\mu \Pi_{\mu\nu} = 0$}.
Here,~$Q$ denotes the four-momentum of the photon.
At finite temperature and/or chemical potential, however, Lorentz invariance is broken and therefore the photon polarization tensor in general depends on~$Q_0$ and~$\vec Q$ separately and can be composed as follows:
\begin{align}
\label{eq:poldecomp}
	\Pi_{\mu\nu} = a A_{\mu\nu} + b B_{\mu\nu} + c\, C_{\mu\nu} + d D_{\mu\nu} \,.
\end{align}
Here, we have introduced four O$(3)$-symmetric projectors~\cite{Gross:1980br}:
\begin{align}
	A_{\mu\nu} &= \delta_{\mu i}\left(\delta_{ij}-\frac{Q_i Q_j}{\vec{Q}\,^2}\right)\delta_{j\nu}\ ,
	\\
	B_{\mu\nu} &= \delta_{\mu\nu} - \frac{Q_\mu Q_\nu}{Q^2} - A_{\mu\nu}\ ,
	\\
	C_{\mu\nu} &= \frac{1}{\sqrt{2}|\vec Q\,|} \left[\left(\delta_{\mu 0} - \frac{Q_\mu Q_0}{Q^2}\right)Q_\nu \right.
	\nn\\
	&\hspace{1.7cm} \left. +\, Q_\mu\left(\delta_{\nu 0}- \frac{Q_\nu Q_0}{Q^2}\right)\right]\ ,
	\\
	D_{\mu\nu} &= \frac{Q_\mu Q_\nu}{Q^2}\,.
\end{align}
The coefficients $a$, $b$, $c$, and $d$ in Eq.~\eqref{eq:poldecomp} are scalar functions which depend on the gauge fixing parameter as well as the momenta $Q_0$ and $\vec Q$. 
Note that the polarization tensor does not depend on the gauge fixing in the large-$\Nf$ limit considered here. 
For a diagrammatic representation of the corresponding one-loop correction to the polarization tensor, see Fig.~\ref{fig:polTensor}.

The polarization tensor can be decomposed in two contributions, the matter part~$\Delta\Pi_{\mu\nu}$ and the vacuum part. 
In the following we shall only consider the matter part and associate its contributions to the coefficients~$a$ and~$b$ with Meissner and Debye screening, respectively. 
In the limit of vanishing temperature and vanishing external momentum, we define the Debye mass~$m_{\rm D}$ and Meissner mass~$m_{\rm M}$ via the following 
{\allowdisplaybreaks 
projection rules:
\begin{align}
	m_{\rm D}^2 &= \lim_{T\to 0}\overset{({\rm st})}{\lim_{Q \to 0}} B_{\nu\mu}\Delta\Pi_{\mu\nu}\,, \\
	m_{\rm M}^2 &= \lim_{T\to 0}\overset{({\rm st})}{\lim_{Q \to 0}} A_{\nu\mu}\Delta\Pi_{\mu\nu}\,,
\end{align}
where} the limit of vanishing external momentum is taken in the form of the static limit.
\begin{figure}[t]
	\includegraphics[]{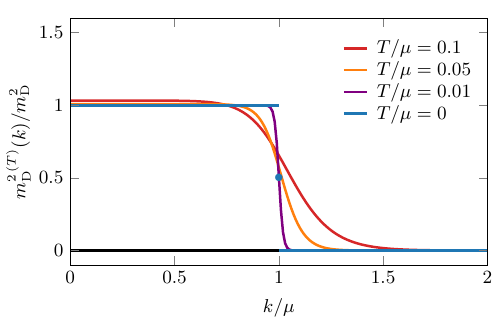}
	\caption{
		Scale-dependent Debye mass as a function of the dimensionless RG-scale $k/\mu$ at zero and various finite temperatures measured in units of IR value of the Debye mass at zero temperature. The zero temperature result has a discontinuity at $k/\mu=1$, see main text for details.}
	\label{fig:DebyeMassRunning}
\end{figure}

The polarization tensor and its static and plasmon limit can be accessed within the fRG approach by applying the expansion defined in Eq.~\eqref{eq:PF_expansion}.
For the Debye mass at finite temperature, we then find the following flow equation:
\begin{widetext}
\begin{align}
\delt m_{\rm D}^{2\,(T)} (k) = \overset{({\rm st})}{\lim_{Q \to 0}} 2 \Nf\, e^2 \int_{\vec p} T \sum_{n \in \mathbb{Z}} \, \delt  \frac{2\,\zeta_k^\tp(\nu_n,\vec{p}^{\,}) \zeta_k(\nu_n + Q_0,-(\vec{p}+\vec Q))}{\hspace{0.2cm}\zeta_k^2(\nu_n,\vec{p}^{\,})\zeta_k^2(\nu_n+Q_0,\vec{p}+\vec Q^{\,})} \, ,
\label{eq:Debye_flow_at_finite_T}
\end{align}	
\end{widetext}
where the definition of the regularized momentum $\zeta_k$ can be found in Eq.~\eqref{eq:zetak}.
Performing now the mathematical operations (derivative with respect to $k$, Matsubara summation, integral over spatial momenta, static limit, and eventually integral over the RG scale~$k$) in the order as given in Eq.~\eqref{eq:Debye_flow_at_finite_T} and using the sharp cutoff introduced in Eq.~\eqref{eq:sharp_cutoff} for convenience, we find that our result for the Debye mass (at $k=0$) agrees with the well-known result in the literature~\cite{Gross:1980br,Toimela:1982ht,Toimela:1984xy}:
\begin{align}
	m_{\rm D}^{2\,(T)} = \Nf\, e^2 \left( \frac{T^2}{3} + \frac{\mu^2}{\pi^2} \right) \, .
	\label{eq:Debye_mass_IR_static}
\end{align}
Note that we have set the Debye mass to zero at the initial RG scale~$\Lambda \gg \mu$ to obtain Eq.~\eqref{eq:Debye_mass_IR_static}. 
For an illustration of the RG flow of the Debye mass, we refer to Fig.~\ref{fig:DebyeMassRunning}. 

By taking the limit~$T\to 0$ in Eq.~\eqref{eq:Debye_flow_at_finite_T} after all mathematical operations have been carried out, we can also derive the flow equation for the Debye mass in the zero-temperature limit. 
It reads
\begin{align}
\label{eq:DebyeT0flow}
\delt m_{\rm D}^{2} (k) = -  k \Nf \frac{e^2  \mu^2}{\pi^2} \delta(\mu -k)\,,
\end{align}
where $m_{\rm D}^{2}=\lim_{T\to 0}m_{\rm D}^{2\,(T)} $ and we have again used the sharp cutoff. 
Setting the Debye mass to zero at the initial RG scale~$\Lambda \gg \mu$, we obtain the following result for the Debye mass as a function of the RG scale~$k$:
\begin{align}
\label{eq:Debye_k}
m_{\rm D}^2(k) = \Nf\frac{e^2 \mu^2}{\pi^2}\theta(\mu - k) \, .
\end{align}
As expected for $k \to 0$, this is in agreement with the zero-temperature limit of the finite-temperature result in Eq.~\eqref{eq:Debye_mass_IR_static}.

The scale dependence  of the zero-temperature Debye mass is depicted in Fig.~\ref{fig:DebyeMassRunning} together with the finite-temperature results.
From this figure, we deduce that the flow of the Debye mass converges in a point-like manner to the zero-temperature result.
Of course, the convergence cannot be uniform since the limiting function is not continuous, at least for the sharp cutoff.\footnote{For a discussion of the Litim regulator, we refer to App.~\ref{app:othercutoff}.}
For smooth regulator functions (e.g., exponential or polynomial regulators), however, we have a continuous limiting function. 
To be more specific, the limiting function is finite for~$k < \mu$, tends to zero continuously for~$k\to \mu$, and remains zero for~$k\geq \mu$. 
In any case, we find that the one-loop result for the Debye mass does not depend on the regulator for~$k\to 0$, as it should be.

In addition to the static limit, we can also consider the plasmon limit. 
The corresponding flow equation is readily obtained from Eq.~\eqref{eq:Debye_flow_at_finite_T} by replacing the static limit with the plasmon limit. 
Again at zero temperature, we find
\begin{align}
\label{eq:m2Dpl}
\delt m^2_{\rm D,\text{(pl)}} = 0
\end{align}
and therefore we also have $m^2_{\rm D,\text{(pl)}} =0$, if we set the initial condition to zero as done above.

As in the case of the Gross-Neveu-Yukawa model, computations directly at zero temperature are potentially problematic. 
To illustrate this again, we simply replace the Matsubara sum in Eq.~\eqref{eq:Debye_flow_at_finite_T} by an integral over the time-like momentum modes~$p_0$.
Performing now all the mathematical operations in the order as given in Eq.~\eqref{eq:Debye_flow_at_finite_T}, we arrive at the following flow equation for the Debye mass at vanishing temperature:
\begin{align}
\label{eq:Debye_flow1}
\partial_t m_{\rm D}^2(k) = 0 \, .
\end{align}
Apparently, this flow equation differs from the one presented in Eq.~\eqref{eq:DebyeT0flow} and therefore also the resulting masses would be different in the IR limit, although we did not change the order of the mathematical operations in Eq.~\eqref{eq:Debye_flow_at_finite_T}. 
Mathematically speaking, this can be traced back to the fact that the derivative with respect to the RG scale~$k$ appearing under the integral over~$p_0$ generates poles in the complex $p_0$-plane which are of order~$n>1$, see our general discussion in Sec.~\ref{sec:Calculation_Subtleties}, in particular Eq.~\eqref{eq:integration_order_p0}.
As also discussed above (see, e.g., Sec.~\ref{subsec:trunc}), loosely speaking, this problem only occurs since it is not possible to solve the Wetterich equation (or, analogously, the continuum path integral) in closed form. 
For example, to obtain the equations for the polarization tensor, we expand the Wetterich equation in terms of the fields by interchanging the loop integration and the derivatives with respect to the fields. 
These field derivatives together with the derivative with respect to the scale~$k$ under the integral over the time-like momentum modes~$p_0$ then lead to incorrect results, see also Eq.~\eqref{eq:integration_order_p0}. 
At least at the one-loop level and provided that we take into account the full momentum dependence of a given correlation function, this can be ``cured" by our hands-on prescription, i.e., by interchanging the integral over~$p_0$ and the derivative with respect to~$k$, as also argued and demonstrated in our study of the Gross-Neveu-Yukawa model in the previous subsection, see also Sec.~\ref{sec:noncomm} for a more general discussion of this prescription.
Indeed, interchanging the integral over~$p_0$ and the derivative with respect to~$k$ but leaving the order of mathematical operations in Eq.~\eqref{eq:Debye_flow_at_finite_T}, we arrive at
\begin{align}
\label{eq:Debye_flow2}
\partial_t m_{\rm D}^2(k) = 2 N_{\rm f} e^2 \int_{\vec p} \partial_t \, \delta \big(\mu-\sqrt{\vec p\,^2(1+r_k)^2}\,\big) \, .
\end{align}
Taking now the derivative with respect to~$k$ and computing the integral by integration by parts indeed yields \mbox{$m_{\rm D}^2 = N_{\rm f} e^2 \mu^2/\pi^2$}, in agreement with the zero-temperature limit of the finite-temperature result in Eq.~\eqref{eq:Debye_mass_IR_static}.

Next, let us briefly discuss the effect of pulling the limit of vanishing external momentum under the loop integral. 
In this case, the integrands of the loop integral as obtained from the plasmon and the static limit agree identically, at zero and finite temperature. 
As a consequence, also these two limits necessarily lead to the same result which is inconsistent with our analysis above.
For example, at zero temperature, we find that the right-hand side of the flow equation given in Eq.~\eqref{eq:Debye_flow_at_finite_T} vanishes identically, irrespective of which of the two limits are pulled under the integral.
Of course, pulling these limits under the loop integral is in general not allowed because of the analytic properties of the two-point function and therefore it does not come as a surprise that we encounter incorrect results in this case.

We close our discussion by briefly commenting on the Meissner mass as well as on the plasmon limit. 
In the static limit, we find that the Meissner mass vanishes at zero and finite temperature for $k\to 0$, again in agreement with the literature~\cite{Gross:1980br,Toimela:1982ht,Toimela:1984xy}. 
To be explicit, in the zero-temperature limit, this mass obeys the following scale dependence:
\begin{align}
m_{\rm M}^2 (k) = - \frac{2\Nf\, e^2}{3 \pi^2}k^2\theta(\mu - k) \,,
\end{align}
where we have again employed the sharp cutoff regulator. 
With respect to the plasmon limit, we would like to add that we obtain the following results in the zero-temperature limit from the projections associated with the Meissner and Debye mass:
\begin{align}
\overset{({\rm pl})}{\lim_{Q \to 0}} B_{\nu\mu}\Delta\Pi_{\mu\nu} = 0 \quad \text{and} \quad 
\overset{({\rm pl})}{\lim_{Q \to 0}} A_{\nu\mu}\Delta\Pi_{\mu\nu} = \frac{\Nf\,e^2 \mu^2}{\pi^2}\,,
\end{align}
see also Eq.~\eqref{eq:m2Dpl}. 
Loosely speaking, the two projections have switched their roles compared to the static limit.
As a consequence, we obtain the same result for the (full) transverse projection (i.e., $P_{\mu\nu}^T = A_{\mu\nu} + B_{\mu\nu}$) of $\Delta\Pi_{\mu\nu}$ from the static and plasmon limit.

\section{Non-commuting operations in functional flows}
\label{sec:noncomm}
After having discussed complications and subtleties encountered in one-loop calculations of correlation functions at zero and finite temperature in the presence of a finite chemical potential, we now would like to consider the situation in functional flows on general grounds. 
To this end, we first analyze the RG flow equation for the effective action at one-loop order.

In Eq.~\eqref{eq:PF_expansion}, we have interchanged summation (associated with an expansion in fields) and integration (as encoded in the trace). 
However, if the integral on the right-hand side of the first line of Eq.~\eqref{eq:Wetterich_truncated} is non-analytic (as it is the case at zero temperature and finite chemical potential), then an interchange of these two operations is in general not allowed. 
Note that this interchange corresponds to an interchange of derivatives with respect to the fields and the loop integration. 
Since the derivative with respect to the RG scale~$k$ under the loop integral increases the order of the poles in the complex~$p_0$-plane, this interchange of differentiation and integration is particularly problematic and leads to incorrect results at zero temperature and finite chemical potential, see Eq.~\eqref{eq:integration_order_p0} for a general discussion. 
Illustrations in terms of concrete loop calculations have been discussed in the previous section. 
In principle, this may be ``cured" by pulling the derivative with respect to~$k$ out of the $p_0$-integral on the right-hand side of Eq.~\eqref{eq:PF_expansion}.
Whereas this is strictly speaking not allowed at this point, it is in fact allowed on the right-hand side of the first line of Eq.~\eqref{eq:Wetterich_truncated}, according to our general discussion in Sec.~\ref{sec:Calculation_Subtleties}. 

Let us now take a more general point of view, for which we assume that the path integral representation of the effective average action is UV-finite, even without the appearance of an RG derivative. 
Pulling the derivative with respect to~$k$ out of the loop integral on the right-hand side of the first line of Eq.~\eqref{eq:Wetterich_truncated} corresponds to directly computing the one-loop approximation of the effective action from the path integral {\it before} taking the derivative with respect to~$k$.
This directly computed one-loop approximation of the effective action can in principle also be expanded in terms of the fields. 
Letting the~$k$-derivative then act on the coefficients of this expansion would yield a series which corresponds to the one in Eq.~\eqref{eq:PF_expansion} but with the loop integral and the derivative with respect to $k$ interchanged. 
However, the so obtained flow equations for the coefficients associated with these two expansions do not necessarily agree since the interchange of the aforementioned operations is not allowed if the associated integrals are non-analytic, which is the case at zero temperature and finite chemical potential.  

Since the computation of the one-loop effective action before taking the derivative with respect to~$k$ should be considered the correct approach, we can derive a simple hands-on prescription to ``cure" the right-hand side of Eq.~\eqref{eq:PF_expansion}. 
More specifically, we propose the prescription of interchanging the RG derivative with the loop integration for Eq.~\eqref{eq:PF_expansion}. 
Strictly speaking, it suffices to interchange the derivative with respect to~$k$ with the integration over the time-like momenta.

Beyond the one-loop approximation, the situation is indeed similar. To explain this, we begin by noting that the Wetterich equation~\eqref{eq:Wetterich_general} can formally be written as follows:
\begin{align}
\label{eq:Wettln}
\dt \Gamma_k[\Phi] 
&= \frac{1}{2}\STr\left[ \tilde{\partial}_t \ln  \left(\Gamma_{k}^{(1,1)}[\Phi] +R_k\right) \right]\,.
\end{align}
Here, we have introduced the generalized scale derivative~$\tilde{\partial}_t$:
\begin{align}
\label{eq:generalized_dt_derivative}
\tilde{\partial}_t=(\partial_t R_k)\frac{\partial}{\partial R_k}\,.
\end{align}
This derivative only acts on the scale dependence of the regulator. 
We add that this derivative operator has only a symbolic meaning here but it can be implicitly defined by the Wetterich equation in specific applications. 
However, concrete representations of~$\tilde{\partial}_t$ in general depend on the loop momentum, at least for the most frequently used classes of regulators in the literature. 
An interchange of this derivative with the loop integration is therefore not allowed.

An expansion of the right-hand side of Eq.~\eqref{eq:Wettln} would in general lead to the same problems as discussed above in case of the one-loop approximation in the presence of a finite chemical potential at zero temperature. 

Following our line of argument in case of the one-loop approximation, we may now be tempted to ``cure" the flow equations for correlation functions resulting from an expansion of Eq.~\eqref{eq:Wettln} in terms of the fields by interchanging the generalized scale derivative
and the supertrace. 
As already mentioned, the operator~$\tilde{\partial}_t$ in general inherits a dependence on the loop momentum from the regulator function and therefore such an interchange is not possible.
Nevertheless, considering spatial regularization schemes which only depend on the spatial momenta, we can interchange the generalized scale derivative and the integral over the time-like momentum modes $p_0$ included the supertrace. 
In case of our one-loop studies, this interchange would have already been sufficient to ``cure" the corresponding flow equations. 
Because of the one-loop structure of the Wetterich equation, we expect that this is also the case for non-perturbative studies. 
In any case, for covariant regulators, this hands-on prescription cannot be applied beyond the one-loop level.

From a formal standpoint, it is important to emphasize that the appearance of the $k$-derivative in the Wetterich equation~\eqref{eq:Wetterich_general} under the loop integral or the generalized scale derivative in Eq.~\eqref{eq:Wettln} is {\it unproblematic} even at zero temperature and finite chemical potential, provided that the Wetterich equation can be solved in closed form, i.e., without relying on, e.g., an expansion in terms of the fields.
Fom a pragmatic standpoint, we finally stress that the aforementioned problems in the evaluation of loop integrals only occur in calculations directly at zero temperature and finite chemical potential and can be circumvented by computing the loop diagrams at finite temperature and then taking the zero-temperature limit afterwards, see Sec.~\ref{sec:Calculation_Subtleties} for a general discussion and the previous section for concrete examples.

\section{Summary}
\label{sec:summary}
In this work, we collected and discussed subtleties which may be encountered in calculations of $n$-point correlation functions of quantum field theories with fermions coupled to a chemical potential. 
In particular, we discussed scenarios in which an interchange of mathematical operations, such as differentiation, integration and limit processes leads to different results for loop integrals. 

For concrete calculations, we employed the fRG approach. 
As for any well-defined framework, the fRG approach is in principle free of mathematical ambiguities since the order of all operations is determined by the Wetterich equation. 
Observables can then be obtained by applying suitable projection rules to its exact solution.
However, exact solutions of this equation exist only for rare, special cases. 
As a result, approximation schemes for the calculation of correlation functions have been developed which may rely on the assumption that at least some of the involved mathematical operations are commutative. 
Although many of these approaches have been applied very successfully to quantum field theories in the vacuum limit, finite external parameters such as temperature or chemical potential introduce several subtleties which must be taken into account in order to obtain correct results for correlation functions. 
As pointed out in detail, calculations at finite chemical potential directly at zero temperature are particularly delicate in this respect. 

In addition to our general discussion of subtleties which arise in finite-temperature and finite-density studies, we demonstrated where such subtleties are encountered in concrete calculations by studying the Gross-Neveu-Yukawa model and QED.
To be specific, we calculated the curvature and the plasmon mass of the boson in the Gross-Neveu-Yukawa model and the photon in QED, both in the limit of many fermion flavors.
To this end, we employed a standard expansion of the Wetterich equation in terms of fields and analyzed the resulting two-point correlation functions associated with the aforementioned particles. 
In accordance with our general considerations, we found that the zero-temperature and finite-temperature results are inconsistent in the sense that the two-point functions calculated directly at zero temperature do not agree with the corresponding finite-temperature results in the zero-temperature limit. 
This is the case for correlation functions evaluated on vanishing external momenta {\it and} finite external momenta. 
We showed that the inconsistency between results for correlation functions obtained at $T=0$ and in the limit~${T \rightarrow 0}$ eventually originates from interchanging derivatives with respect to fields with the loop integration on the right-hand side of the Wetterich equation. 
In conjunction with the presence of a derivative with respect to the RG scale under the loop integral, this is not allowed in calculations directly at zero temperature for values of the chemical potential that exceed the Silver-Blaze threshold because of the non-analytic behavior of the loop diagrams in this regime. 
From our analysis, however, we deduced a prescription which allows to compute correlation functions directly at zero temperature such that they are consistent with those obtained from taking the zero-temperature limit of a finite-temperature calculation. 
We also showed that finite-temperature calculations of correlation functions via the aforementioned interchange of derivatives with respect to fields and the loop integration are unambiguous, even at finite chemical potential. 

We close by noting that, although we focussed on diagrams with a one-loop structure, as only such diagrams appear within the fRG framework, 
our general considerations can be carried over to computations of loop diagrams of higher order as encountered in other approaches.
In any case, our analysis makes clear that the computation of correlation functions in the presence of a finite chemical potential at zero temperature requires great care to ensure consistency with corresponding finite-temperature calculations. 

{\it Acknowledgments.--} We thank Tyler Gorda, Jan M. Pawlowski, and Fabian Rennecke for discussions and comments on the manuscript. 
Moreover, we thank Adrian K\"onigstein for discussions. 
As members of the fQCD collaboration~\cite{fQCD}, the authors also would like to thank the members of this collaboration for discussions. 
This work is supported in part by the Deutsche Forschungsgemeinschaft (DFG, German Research Foundation) through the Collaborative Research Center CRC~1245 ``Nuclei: From Fundamental Interactions to Structure and Stars" -- project number 279384907 -- SFB 1245 and the Collaborative Research Center TransRegio CRC-TR 211 ``Strong-interaction matter under extreme conditions" -- project number 315477589 -- TRR 211.

\appendix
\section{Matsubara sum and differentiation}
\label{sec:app1}
We consider a function $f(z,y)$, which is analytic in $z \in \mathbb{C} \setminus P$, where $P$ denotes a finite set of isolated points. 
Suppose that the Matsubara series of $f(\nu_m(\beta),y)$ exists for all $y \in \mathbb{R}$ and $\beta >0$, then
\begin{align}
\label{eq:commutation_relation_2}
\frac{1}{\beta} \sum_{m \in \mathbb{Z}} \partial_y f(\nu_m(\beta),y) = \frac{\d}{\d y} \frac{1}{\beta} \sum_{m \in \mathbb{Z}} f(\nu_m(\beta),y)\, .
\end{align}
In words, the Matsubara summation commutes with the derivative with respect to a $\beta$-independent variable. 
Note that this is in general a non-trivial statement since infinite sums do not need to preserve the linearity of the operator applied to them. 
The statement above relies on the fact that the infinite sum of contributions from different Matsubara frequencies $\nu_m$ can be turned into a finite sum of residues by means of the Matsubara formalism: 
\begin{align}
 \frac{1}{\beta} \sum_{m \in \mathbb{Z}}& \partial_y f(\nu_m(\beta),y) \nn\\
&  = - \iu \sum_{z^\ast \in P} \Res{\partial_y f(\bfdot, y)\ \frac{1}{\e^{-\iu \beta \bfdot}+1}, z^\ast} \nn\\
&  = - \iu \sum_{z^\ast \in P} \Res{\partial_y\left[ f(\bfdot, y)\ \frac{1}{\e^{-\iu \beta \bfdot}+1}\right], z^\ast} \nn \\
  \overset{\eqref{eq:commutation_relation_1}}&{=} - \iu \sum_{z^\ast \in P} \frac{\d}{\d y} \Res{f(\bfdot, y)\ \frac{1}{\e^{-\iu \beta \bfdot}+1}, z^\ast} \nn\\
  \overset{|P|<\infty}&{=} - \iu \frac{\d}{\d y} \sum_{z^\ast \in P} \Res{f(\bfdot, y)\ \frac{1}{\e^{-\iu \beta \bfdot}+1}, z^\ast} \nn \\
&  =\frac{\d}{\d y} \frac{1}{\beta} \sum_{m \in \mathbb{Z}} f(\nu_m(\beta),y)\, .
\end{align}
Here, we have used that
\begin{align}
\label{eq:commutation_relation_1}
\Res{\partial_y f(\bfdot, y), g(y)} = \frac{\d}{\d y} \Res{f(\bfdot, y), g(y)}\,,
\end{align}
where the real-valued variable $y$ may refer to a multitude of quantities related to the physical system under consideration, such as the RG scale, a mass, or an external momentum.
Note that we restrict ourselves here to functions $f(z,y)$ of the form~\eqref{eq:generic_meromorphic_function}, which are analytic in $z \in \mathbb{C}$ except for the isolated point $z^\ast= g(y)$. 
Relation~\eqref{eq:commutation_relation_1} can then be proven as 
{\allowdisplaybreaks follows:
\begin{widetext}
\begin{align}
\Res{\partial_y f(\bfdot,y ), g(y)} &=
\Res{\frac{h^{(0,1)}(\bfdot,y)}{(\bfdot - g(y))^n} + n \left(\frac{\d}{\d y} g(y)\right) \frac{h(\bfdot,y)}{(\bfdot - g(y))^{n+1}}, g(y) }\nn \\
&= \frac{1}{n!} \lim_{z\rightarrow g(y)} \frac{\partial^n}{\partial z^n}\left[(z-g(y))\ h^{(0,1)}(z,y) + n \left(\frac{\d}{\d y} g(y)\right) h(z,y) \right]\nn \\
&= \frac{1}{n!} \lim_{z\rightarrow g(y)} \left[ \sum_{j=0}^n \binom{n}{j} \left(\frac{\partial^j}{\partial z^j}(z-g(y))\right)\ h^{(n-j,1)}(z,y) + n  \left(\frac{\d}{\d y} g(y)\right) h^{(n,0)}(z,y) \right]\nn \\
&=  \frac{1}{n!} \lim_{z\rightarrow g(y)} \left[ (z-g(y))\ h^{(n,1)}(z,y) +n\ h^{(n-1,1)}(z,y) +  n \left(\frac{\d}{\d y} g(y)\right) h^{(n,0)}(z,y) \right]\nn \\
&=  \frac{1}{(n-1)!} \left[ h^{(n-1,1)}(g(y),y) +   \left(\frac{\d}{\d y} g(y)\right) h^{(n,0)}(g(y),y) \right]\nn \\
&= \frac{1}{(n-1)!}\ \frac{\d}{\d y} h^{(n-1,0)}(g(y),y) \nn \\
&= \frac{\d}{\d y}\ \frac{1}{(n-1)!} \lim_{z \rightarrow g(y)} \frac{\partial^{n-1}}{\partial z^{n-1}} h(z,y)\nn \\ 
&= \frac{\d}{\d y}\ \frac{1}{(n-1)!} \lim_{z \rightarrow g(y)} \frac{\partial^{n-1}}{\partial z^{n-1}} \left[(z-g(y))^n\ f_k(z,y) \right]\nn \\
&= \frac{\d}{\d y} \Res{f(\bfdot,y), g(y)}\ .
\end{align}
\end{widetext}
}

\section{Computation of the effective potential}
\label{app: effective potential}
Here, we provide details underlying our computation of the effective potential of the Gross-Neveu-Yukawa model at zero temperature. 
More specifically, we would like to comment on the integration over the time-like momentum modes as done in Eq.~\eqref{eq:effective_potential}. 
For convenience, we will use the notation
\begin{align}
x_k^2 = \vec{p}^{\,2}(1+r_k)^2 + h^2 \sigma^2\,,
\end{align}
such that the entire dependence of the logarithmic integrand on spatial momenta as well as field degrees of freedom is kept implicit. 
In Eq.~\eqref{eq:effective_potential}, the integration with respect to the time-like momentum modes can be performed by integration by parts and the Cauchy residue theorem. To be more concrete, assuming that $\mu \geq 0$, we obtain
{\allowdisplaybreaks
\begin{align}
\label{eq: effective potential calculation}
\int_{\mathbb{R}} \frac{\d p_0}{2\pi} &\left.\ln( \left(p_0 + \iu \mu \right)^2 + x_{k^\prime}^2 )\right|^{k^\prime = k}_{k^\prime = \Lambda_0} \nn\\
&= -2 \int_{\mathbb{R}} \frac{\d p_0}{2\pi} \frac{p_0 \left(p_0 + \iu \mu\right)}{\left(p_0 + \iu \mu \right)^2 + x_{k^\prime}^2 }\Bigg|^{k^\prime = k}_{k^\prime = \Lambda_0} \nn\\[2mm]
&=\left.\Big[(|x_{k^\prime}|-\mu)\ \theta\left(|x_{k^\prime}|-\mu \right)\Big]\right|^{k^\prime = k}_{k^\prime = \Lambda_0}\ .
\end{align}
We would like} to remark that the consistent implementation of an IR and UV regularization, as provided by the fRG approach, naturally leads to a vanishing of boundary terms here.
Further note that, through integration by parts, we have mapped the integral over a logarithmic function onto an integral over a function which has a simple complex pole.
From our analysis in Sec.~\ref{subsec:2} it then follows that the result \eqref{eq: effective potential calculation} agrees identically with the zero-temperature limit of a corresponding finite-temperature calculation. 
Indeed, we have
{\allowdisplaybreaks
\begin{align}
&\lim_{\beta \to \infty} \frac{1}{\beta} \sum_{n \in \mathbb{Z}} \left.\ln( \left(\nu_n(\beta) + \iu \mu \right)^2 + x_{k^\prime}^2 )\right|^{k^\prime = k}_{k^\prime = \Lambda_0} \nn\\
&= \lim_{\beta \to \infty} \left.\left[|x_{k^\prime}| + \frac{1}{\beta} \sum_{\pm} \ln\left(1 + \e^{-\beta(|x_{k^\prime}|\pm \mu)} \right)\right]\right|^{k^\prime = k}_{k^\prime = \Lambda_0}\nn\\
&= \left.\Big[ |x_{k^\prime}| + (\mu - |x_{k^\prime}|)\ \theta(\mu - |x_{k^\prime}|) \Big]\right|^{k^\prime = k}_{k^\prime = \Lambda_0}\nn\\
&= \left.\Big[  (|x_{k^\prime}|- \mu )\ \theta(|x_{k^\prime}|- \mu) \Big]\right|^{k^\prime = k}_{k^\prime = \Lambda_0}\,.
\end{align}
This is in accordance with the standard result in the literature, see, e.g., Ref.~\cite{Hong:2000rk}.}
\begin{figure}[t]
	\includegraphics[]{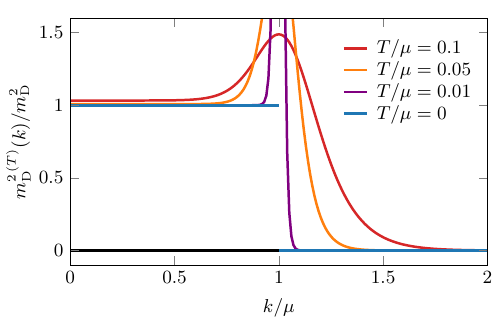}
	\caption{Debye mass of the photon as a function of the dimensionless RG scale $k$ at zero and finite temperature for a given fixed value of the chemical potential $\mu$ as obtained from the Litim regulator. The finite-temperature results indicate that, for this regulator, the RG flow of the mass receives a contribution from a Dirac delta distribution at~$k=\mu$ in the limit of vanishing temperature which is indeed confirmed by our analytic results, see Eq.~\eqref{eq:Debye_k_Litim}.}
	\label{fig:DebyeMassRunningLitim}
\end{figure}
\section{Litim regulator}
\label{app:othercutoff}
In the main text, we used a sharp cutoff in all explicit calculations of correlation functions.  
However, the complications and subtleties which are present in finite-density calculations of correlation functions at zero temperature do not originate from the use of this particular cutoff.\footnote{The sharp cutoff regulator is known to generate ambiguities in studies relying on a derivative expansion of the effective action. In our present work, we did not employ a derivative expansion at all but always considered the full momentum dependence of correlation functions.} 
This already becomes clear from our general discussion. 
To make this explicit, we present a computation of the Debye mass of the photon with the three-dimensional version of the Litim regulator~\cite{Braun:2003ii,Litim:2006ag,Blaizot:2006rj}:
\begin{align}
\label{eq:Litim}
r(x) = \left(\frac{1}{\sqrt{x}}-1 \right) \theta(1-x) \, .
\end{align}
Within the loop integral, the non-analytic behavior of the Litim regulator becomes manifest at momenta which are small compared to the RG scale:
\begin{align}
\vec{p}^{\,2} \left(1+ r\left( \frac{\vec{p}^{\,2} }{k^2}\right)\right)^2 = \max\left(|\vec{p}^{\,}|,k \right)^2\, .
\end{align}
For the computation of the Debye mass, we follow Sec.~\ref{subsec:photon_curvature_masses}. 
To be specific, from Eq.~\eqref{eq:Debye_flow_at_finite_T}, which is valid for general three-dimensional regulator functions, we obtain
\begin{align}
	m_{\rm D}^2(k) 
	= 2 N_{\rm f} e^2 \int_{\vec p} \delta\big(\mu-\sqrt{\vec p\,^2(1+r_{k'})^2}\,\big)\Big|^{k'=k}_{k'=\Lambda}\,.
\label{eq:mD2genreg}
\end{align}
Here, the integral with respect to the time-like momentum modes has been performed before the derivative with respect to the RG scale~$k$. 
According to our discussion in Sec.~\ref{subsec:photon_curvature_masses}, this ensures consistency with the zero-temperature limit of a corresponding finite-temperature calculation.  

In any case, by plugging the definition~\eqref{eq:Litim} of the Litim regulator into Eq.~\eqref{eq:mD2genreg}, we obtain
{\allowdisplaybreaks
\begin{align}
	m_{\rm D}^2(k) 
	 = 2 N_{\rm f} e^2\left[  \int_{|\vec p\,|\leq k} \delta(\mu\!-\! k) \!+\! \int_{|\vec p\,| > k} \delta(\mu\! -\! |\vec p\,|) \right]\,.\nn
\end{align}
Evaluating} the integrals over the spatial momenta, we finally find
\begin{align}
	m_{\rm D}^2(k)  =  \frac{N_{\rm f}e^2 \mu^2}{\pi^2}\theta(\mu - k)+ \frac{N_{\rm f}e^2}{3 \pi^2} k^3\ \delta(\mu-k) \, . 
	\label{eq:Debye_k_Litim}
\end{align}
Note that two terms appear on the right-hand side. 
The first term associated with the Heaviside step function agrees identically with the scale dependence of the Debye mass as obtained from our calculation with the sharp cutoff, see Eq.~\eqref{eq:Debye_k}.
The emergence of the second term in Eq.~\eqref{eq:Debye_k_Litim} associated with a Dirac delta distribution can be traced back to the Litim regulator being non-analytic.
We emphasize that this term does not result from an interchange of mathematical operations that would not be valid. 
In this respect, this Dirac delta distribution should not be confused with the Dirac delta distributions which appear in our results for, e.g., the boson curvature mass in the Gross-Neveu-Yukawa model, see Eq.~\eqref{eq:curvature_mass_MF_T=0}.

Considering the RG flow of the Debye mass at finite temperature, the emergence of the contribution associated with the Dirac delta distribution in Eq.~\eqref{eq:Debye_k_Litim} can be observed by decreasing temperature while keeping the chemical potential fixed, see Fig.~\ref{fig:DebyeMassRunningLitim}.
In any case, as can be deduced from Eq.~\eqref{eq:Debye_k_Litim}, this contribution vanishes in the limit~$k\to 0$ and the result for the Debye mass then agrees identically with the well-known result from the literature, see Eq.~\eqref{eq:Debye_mass_IR_static}. 
\bibliography{qcd}
\end{document}